\documentclass[iop]{emulateapj}
\usepackage{natbib}
\usepackage{multirow}
\usepackage{latexsym}
\usepackage{amsmath}
\usepackage[breaklinks]{hyperref}
\usepackage{subfigure}
\newcommand{\hncone}{\mbox{\rmfamily HNC}\,{(1--0)}\,}

\newcommand{\hcnone}{\mbox{\rmfamily HCN}\,{(1--0)}\,}

\newcommand{\nthp}{\mbox{\rmfamily N$_{2}$H$^+$}}
\newcommand{\nthpone}{\mbox{\rmfamily N$_{2}$H$^+$}\,{(1--0)}\,}
\newcommand{\hcopone}{\mbox{\rmfamily HCO}$^+$\,{(1--0)}\,}
\newcommand{\hcop}{\mbox{\rmfamily HCO}$^+$}
\newcommand{\htcop}{\mbox{\rmfamily H$^{13}$CO}$^+$}
\newcommand{\htcopone}{\mbox{\rmfamily H$^{13}$CO}$^+$\,{(1--0)}\,}

\newcommand{\hii}{\mbox{ H\,{\sc ii}}}

\shorttitle{Chemistry in Star Formation}
\shortauthors{Hoq et al.}

\begin{document}

\title{Chemical Evolution in High-Mass Star-Forming Regions: Results from the MALT90 Survey}

\author{Sadia Hoq\altaffilmark{1}, James M. Jackson\altaffilmark{1}, Jonathan B. Foster\altaffilmark{1,2}, Patricio Sanhueza\altaffilmark{1}, Andr\'es Guzm\'an\altaffilmark{3}, J. Scott Whitaker\altaffilmark{4}, Christopher Claysmith\altaffilmark{1}, Jill M. Rathborne\altaffilmark{5}, Tatiana Vasyunina\altaffilmark{6}, Anton Vasyunin\altaffilmark{6,7}} 

\altaffiltext{1}{Institute for Astrophysical Research, Boston University, Boston, MA 02215; shoq@bu.edu, jackson@bu.edu,patricio@bu.edu, claysmit@bu.edu}

\altaffiltext{2}{Yale Center for Astronomy and Astrophysics, Yale University, New Haven, CT 06511; jonathan.b.foster@yale.edu}

\altaffiltext{3}{Harvard-Smithsonian Center for Astrophysics, Cambridge, MA 02138; aguzmanf@cfa.harvard.edu}
\altaffiltext{4}{Physics Department, Boston University, Boston, MA 02215; scott@bu.edu}
\altaffiltext{5}{Australia Telescope National Facility, CSIRO Astronomy and Space Science, Epping, NSW, Australia; rathborne@csiro.au}

\altaffiltext{6}{Department of Chemistry, University of Virginia, Charlottesville, VA 22904; tv3h@virginia.edu, aiv3f@virginia.edu}
\altaffiltext{7}{Visiting Scientist, Ural Federal University, Ekaterinburg, Russia}

\begin{abstract}

The chemical changes of high-mass star-forming regions provide a potential method for classifying their evolutionary stages and, ultimately, ages.  In this study, we search for correlations between molecular abundances and the evolutionary stages of dense molecular clumps associated with high-mass star formation.  We use the molecular line maps from Year 1 of the Millimetre Astronomy Legacy Team 90 GHz (MALT90) Survey. The survey mapped several hundred individual star-forming clumps chosen from the ATLASGAL survey to span the complete range of evolution, from prestellar to protostellar to \hii\ regions.  The evolutionary stage of each clump is classified using the {\it Spitzer} GLIMPSE/MIPSGAL mid-IR surveys.  Where possible, we determine the dust temperatures and H$_2$ column densities for each clump from {\it Herschel} Hi-GAL continuum data. From MALT90 data, we measure the integrated intensities of the \nthp, \hcop, HCN and HNC (1-0) lines, and derive the column densities and abundances of \nthp\ and \hcop.  The {\it Herschel} dust temperatures increase as a function of the IR-based {\it Spitzer} evolutionary classification scheme, with the youngest clumps being the coldest, which gives confidence that this classification method provides a reliable way to assign evolutionary stages to clumps.  Both \nthp\ and \hcop\ abundances increase as a function of evolutionary stage, whereas the \nthp\ (1-0) to \hcop\ (1-0) integrated intensity ratios show no discernable trend.  The HCN (1-0) to HNC(1-0) integrated intensity ratios show marginal evidence of an increase as the clumps evolve.

\end{abstract}

\section{Introduction}
\label{Introduction}

The chemical composition of molecular gas undergoing star formation is predicted to evolve due to the physical changes that occur during the star formation process.  As material collapses to form a star, temperatures and densities rise, leading to the production and destruction of different molecular species.  Several theoretical models predict changing molecular abundances of low-mass star-forming molecular cores as they evolve from the prestellar to protostellar phase \citep[e.g.,][]{Bergin1997, Bergin2002, Lee2004, Aikawa2001, Aikawa2005, Aikawa2008}, and have since been substantiated by many low-mass observational studies \citep[e.g.,][]{Caselli1999, Caselli2002, Tafalla2002, Bergin2007, Bergin2002, Tafalla2004, Jorgensen2004}.  

Compared to the number of studies of low-mass star-forming regions, there have been far fewer studies of chemistry in high-mass star-forming regions.  This dearth of studies is partly due to the accelerated timescales of the formation of massive stars and their propensity to form not in isolated environments but rather within clusters.  High-mass stars (M $\ge$ 8 M$_{\odot}$) form within cluster-forming clumps (size $\sim$0.5 pc) \citep[e.g.,][]{Reiter2011} associated with infrared dark clouds \citep{Perault1996, Egan1998, Rathborne2006, Simon2006}.  Several studies of their formation have focused on high-mass protostellar objects (HMPOs) or hot cores \citep[e.g.,][]{Blake1987, Pirogov2003, Nomura2004, Beuther2009, Purcell2009}.  Recently, more studies have focused on the chemistry in earlier evolutionary stages found in IRDCs \citep{Sakai2008, Sakai2010, Sakai2012, Sanhueza2012, Sanhueza2013, Vasyunina2011, Vasyunina2012}.  

 In chemical models of low-mass star formation, a relative enhancement of \nthp\ and a depletion of \hcop\ abundances are expected in the cold, prestellar phase \citep{Bergin1997, Lee2003, Lee2004, Bergin2007}, and as the core warms, the \hcop\ abundance increases, while the \nthp\ abundance drops.  The behaviors of \nthp\ and \hcop\ abundances in molecular cores are largely affected by the presence or absence of CO, a parent molecule of \hcop\ and a major destroyer of gas-phase \nthp.  The CO abundance is significantly depleted in cold, prestellar cores as it freezes onto the surface of dust grains \citep[e.g.,][]{Caselli1999}, and rises as CO is released from the surface of dust grains as the cores warm and heat their surrounding environments.  In the case of HCN and HNC, observational studies \citep[e.g.,][]{Goldsmith1981, Goldsmith1986, Hirota1998} of star-forming cores have shown that their abundance ratio changes as a function of temperature and evolution.

The present study uses observations of the largest sample to date of molecular clumps that span the range of evolution of high-mass star formation from prestellar clumps to \hii\ regions.  As part of the MALT90 Survey \citep[Jackson et al. 2013, in preparation]{Foster2011}, which uses the ATNF Mopra Telescope, several molecular transitions toward each clump were observed and its evolutionary stage was classified into three main categories of prestellar, protostellar, or \hii/PDR region based on {\it Spitzer} mid-IR data (see Section \ref{Classifications}).  We use molecular line observations of \nthpone, \hcopone, \htcop\ (1-0), \hcnone, and \hncone\ to determine whether the chemical changes with evolution predicted by previous studies are observed for the present sample of massive clumps.  These lines were chosen because they are unambiguous tracers of dense molecular clumps due to their high critical densities (n$_{crit}$ $\ge$ 10$^5$ cm$^{-3}$).  We also determine if a strong correlation between molecular abundances and evolutionary stages of star-forming regions can be established, which could be used as a chemical clock to find the evolutionary stage of a star-forming clump or core based on its molecular abundance information. 

Based on the observations of the MALT90 sample, we aim to answer several questions about the chemistry of high-mass star formation.  These questions and the techniques we will employ to answer them are as follows:

%1. Does the MALT90 Survey probe a large sample of high-mass star-forming regions spanning the entire range of star formation evolution?  To establish whether the MALT90 targets fulfill these criteria, we must determine the mass and classify the evolutionary stage of each clump based on its {\it Herschel} submm and {\it Spitzer} mid-IR emission.  Typical masses of clumps which can form a massive star are $\sim$10$^2$--10$^3$ M$_{\odot}$ \citep{Zinnecker2007}.

1. Is the evolutionary classification of the MALT90 sample based on the examination of {\it Spitzer} mid-IR emission reliable and consistent?  As this process is subjective, it may sometimes fail.  Moreover, since distant sources can sometimes be obscured by line-of-sight material, their classifications are less reliable.  Therefore, we test the {\it Spitzer} IR classifications by deriving dust temperatures from {\it Herschel} Hi-GAL Survey data \citep{Molinari2010}.  As a cold prestellar clump evolves into a protostar and eventually into an \hii\ region, it heats the surrounding dust.  Therefore, we expect a monotonic increase in the dust temperature as a clump evolves.  Hence, if the sources in the prestellar, protostellar, and \hii/PDR categories, respectively, show increasing dust temperatures, we can gain confidence that our classification method is reliable.

2. Is there a relative increase in the abundance of HCN compared to that of HNC as a function of evolution, as found by previous observational studies \citep[e.g.,][]{Hirota1998}?  We measure the integrated intensities of the \hcnone and \hncone molecular lines from the MALT90 sample and determine whether the integrated intensity ratios increase as a function of the evolutionary stages of the clumps.

3. Do the molecular observations of MALT90 agree or disagree with the predicted evolution of \nthp\ and \hcop\ abundances from low-mass star formation chemical models?  To answer this question, we measure the integrated intensities and abundances of the \nthpone\ and \hcopone\ molecular lines in the sample and determine how these properties change as a function of the evolutionary stage of the sources.

4. Can the chemical evolution in high-mass star-forming regions be used as a chemical clock?  If a distinct correlation can be established between the evolutionary stage and measured molecular properties of integrated intensities and abundances of the sample, then this relationship can potentially be used to determine the evolutionary phase, and hence relative age, of other star-forming regions based solely on their molecular line emission.

\section{Observations}
\label{Observations}

Observations were taken with the 22-m single dish Australia Telescope National Facility (ATNF) Mopra Telescope as part of Year 1 of the Millimetre Astronomy Legacy Team 90 GHz (MALT90) Survey.  Target sources were chosen from the APEX Telescope Large Area Survey of the Galaxy (ATLASGAL) Catalog \citep{Schuller2009, Contreras2013} of compact 870 $\mu$m continuum sources, with a flux limit of 0.25 Jy. ATLASGAL sources span a wide range of evolutionary stages and had high detection rates in the MALT90 Pilot Survey \citep[see][]{Foster2011}.  Observations were taken from June to September 2010 of 499 high-mass star-forming regions.  Mopra's broadband spectrometer, MOPS, was used to map each source in 16 lines simultaneously at frequencies between 86--93 GHz, with a velocity resolution of $\sim$0.11 km s$^{-1}$.  The Mopra beamwidth is 38'' at 86 GHz \citep{Ladd2005}.  

The data were reduced using the LIVEDATA and GRIDZILLA packages from ATNF.  The intensities are given in antenna temperature corrected for atmospheric opacity (T$^{*}_{\rm A}$), which can be converted to main beam brightness temperature by dividing by the main beam efficiency, $\eta_{mb}$ \citep[$\eta_{mb}$ = 0.49 at 86 GHz][]{Ladd2005}.  Each source was mapped over an area of 4' $\times$ 4'.  Off positions were automatically chosen one degree in Galactic latitude ($b$) away from the source.  Off positions for sources at positive latitudes were located at $b$+1$^{\circ}$, and off positions for sources at negative latitudes were located at $b$-1$^{\circ}$.

In the present study, the data were averaged over a circular area positioned on the center of each map with a diameter roughly equal to the Mopra beamwidth.   The average noise in T$^{*}_{\rm A}$ (T$_{\rm rms}$) for the average spectrum in this central beam is typically $\sim$0.27 K per channel.  We focus on 5 molecular lines from the MALT90 Survey-- \nthpone, \hcopone, \htcopone, \hcnone, and \hncone.  The MALT90 sources were classified into three different evolutionary stages of quiescent, protostellar, and \hii/PDR based on their {\it Spitzer} mid-IR emission (see Section \ref{Classifications}).  Of the 499 sources observed, only 333 were confidently classified into one of the three stages.

We only report findings of these 333 sources with evolutionary classifications.  The coordinates and evolutionary classifications of the 333 sources are listed in Table \ref{continuumtablesample}.  We refer to the structures defined by the Mopra beam as clumps.

The MALT90 data, which include (l,b,v) data cubes and (l,b) moment and error maps, can be accessed at \url{http://malt90.bu.edu}, and \url{http://atoa.atnf.csiro.au/}.

%\begin{center}
\begin{deluxetable*}{cccccc}
\tabletypesize{\scriptsize}
\tablecolumns{6}
%\tablecaption{Parameters Derived from {\it Herschel} Hi-GAL Continuum Data\label{continuumtablesample}} %for a separate table of coords and herschel
\tablecaption{MALT90 Clump Parameters\label{continuumtablesample}}  %for combined coords and herschel info
%\rotate
\tablewidth{0pt}
\tablehead{
 \colhead{Source} &
 \colhead{RA (J2000)} &
 \colhead{DEC (J2000)} &
 \colhead{{\it Spitzer}}&
 \colhead{Temperature} &
 \colhead{H$_2$ Column Density} \\
 \colhead{     } &
 \colhead{     } &
 \colhead{     } &
 \colhead{Classification} &
 \colhead{ (K)} &
 \colhead{($\times$10$^{22}$ cm$^{-2}$)}
}
\startdata
G000.253+00.016  & 17:46:09.6 & -28:42:42.6  & Quiescent &   18.3$^{+2.8}_{-1.8}$   &   31.1$\pm$7.8  \\ 
G000.414+00.045  & 17:46:25.7 & -28:33:33.2  & Quiescent &   16.9$^{+2.5}_{-1.7}$   &   20.8$\pm$5.2   \\ 
G003.033+00.405  & 17:51:07.1 & -26:07:45.7  & Quiescent &   13.9$^{+2.1}_{-1.4}$   &   2.4$\pm$0.6   \\ 
G003.089+00.164  & 17:52:10.1 & -26:12:15.3  & Quiescent &   15.6$^{+2.3}_{-1.6}$   &   3.9$\pm$1.0   \\ 
\enddata
%\tablenotetext{a}{The mass derived for source G000.253+00.016 in the present study is smaller than that found by \citet{Longmore2012} because they find the mass for the entire cloud.}
\tablecomments{This table is available in its entirety in a machine-readable form in the online journal. A portion is shown here for guidance regarding its form and content.}
\end{deluxetable*}
%\end{center}

\subsection{Mid-IR Classification of Sample}
\label{Classifications}
\begin{figure*}
%\begin{sideways}
\begin{center}
\epsscale{0.7}
\plotone{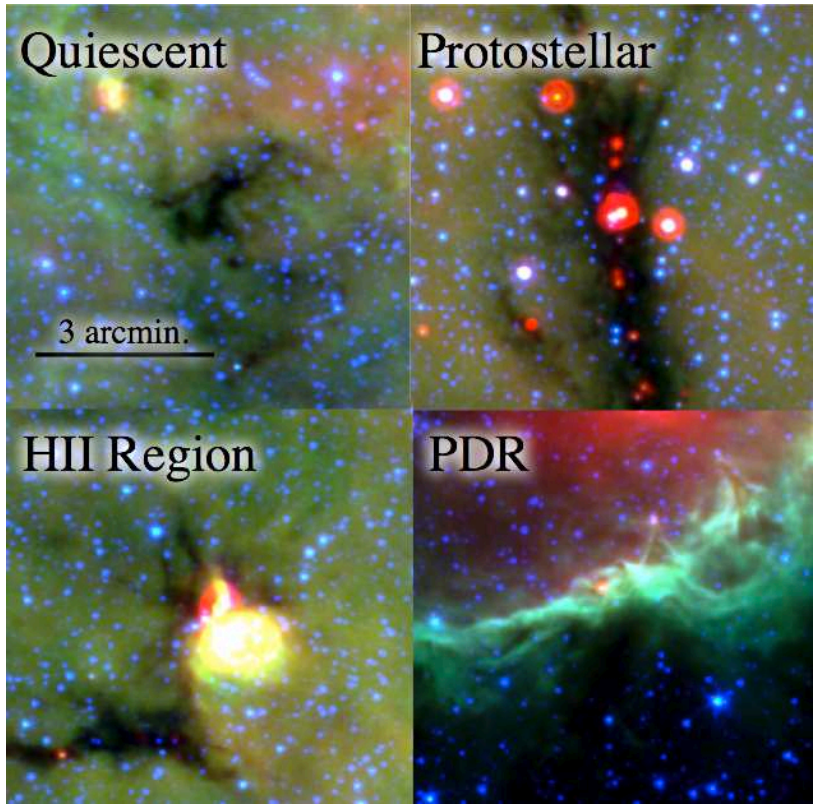}
\caption{Examples of IR clump classification.  Images are GLIMPSE+MIPSGAL (blue: 3.6 $\mu$m, green: 8.0 $\mu$m, red: 24 $\mu$m)}
\label{coretypesnew}
\end{center}
%\end{sideways}
\end{figure*}

Based on their mid-IR characteristics in the {\it Spitzer} GLIMPSE and MIPSGAL surveys \citep{Churchwell2009, Carey2009}, the MALT90 clumps were classified into three different evolutionary stages---quiescent (prestellar), protostellar, and \hii/PDR.  Prestellar clumps are cold, dense regions of gas and dust that show no obvious signs of embedded stars or protostars.  Sources in the MALT90 sample that appeared dark in the 3.6 $\mu$m, 4.5 $\mu$m, 8 $\mu$m, and 24 $\mu$m bands were classified as quiescent.  Clumps that have unresolved emission in the MIPS 24 $\mu$m band or are associated with extended 4.5 $\mu$m emission, so-called ``extended green objects'' or ``green fuzzies'' \citep[see][]{Cyganowski2008, Chambers2009}, were categorized as protostellar.  The unresolved 24 $\mu$m emission of a protostellar source originates from locally heated dust surrounding the protostar, whereas the extended 4.5 $\mu$m emission is likely associated with shocks generated by molecular outflows.  \hii\ regions and photodissociation regions (PDRs), which are the interfaces between \hii\ regions and surrounding molecular material, show extended emission at 8 $\mu$m, due to flourescently-excited polycyclic aromatic hydrocarbon (PAH) emission contained in the 8 $\mu$m bandpass.  Figure~\ref{coretypesnew} shows an example of a source at each evolutionary stage.  The prestellar source appears as a dark extinction feature superimposed against a brighter background.  Embedded protostars correspond to the 24 $\mu$m point sources.  The \hii\ region emits extensively in the {\it Spitzer} bands, and the PDR corresponds to extended PAH emisson at 8 $\mu$m.

%Figure 1

While classifying the sample, we only considered emission within a circular area of diameter 38'' centered on each clump, roughly equal to the angular resolution of the main beam of the Mopra telescope at 90 GHz.  Each source was classified by eye by several individuals to limit bias and erroneous classification.  A source where the classification was difficult or ambiguous (for example, when two different individuals did not agree on the classification) was not included in further analysis; only sources that were the most reliably classified were used.  333 were reliably categorized into one of the three classifications---59 quiescent, 95 prostellar, and  179 \hii/PDR clumps.

\section{Results}
\label{Results}
\subsection{Dust Temperatures and H$_2$ Column Densities}
\label{dust temperatures and h2 column densities}

Dust temperatures were derived from \emph{Herschel} Hi-GAL Survey data by using the emission detected at 160, 250, 350, and 500 $\mu$m.  Following the analysis of \citet{Battersby2011}, we exclude the flux measured at 70 $\mu$m using the blue filter of the PACS instrument because a large fraction of this flux is likely dominated by very small grain emission. Furthermore, at 70 $\mu$m, many IRDCs from the MALT90 sample are seen in absorption against the Galactic plane diffuse FIR background.  Hence, the 70 $\mu$m emission is unreliable as an indicator of dust temperature.

To derive the fluxes associated with each MALT90 source, we first subtract an extended component from the \emph{Herschel} maps that represents the contribution from the foreground and background diffuse emission.  From the subtracted map, the average flux was measured within a 38'' FWHM Gaussian beam centered in the position of each MALT90 source.  

To derive temperatures and H$_2$ column densities, we fit the fluxes measured at each band by using a single temperature emission model given by
\begin{equation}
I_\nu=B_\nu(T_{\rm d})\left(1-\exp\left(- \kappa_\nu\mu {\rm m}_{\rm H}{\rm N}_{\rm H_2} \right)\right),
\end{equation}
where $I_\nu$ is the intensity at frequency $\nu$, $B_\nu(T_{\rm d})$ is the Planck function at dust temperature $T_{\rm d}$, $\kappa_\nu$ is the dust opacity, $\mu$ corresponds to the mean molecular weight, which is assumed to equal 2.3, m$_{\rm H}$ is the mass of one hydrogen atom, and ${\rm N}_{\rm H_2}$~is the H$_2$~column density.  We defer the detailed description of the extended emission subtraction algorithm and the fitting procedure to an upcoming paper (A. Guzm\'an et al., in preparation)

The dust absorption coefficient $\kappa_\nu$ was derived from the tables of \citet{Ossenkopf1994} assuming dust coagulation at a density of $10^5$ cm$^{-3}$, thin ice mantle coating, and a gas-to-dust mass ratio equal to 100\footnotemark[1].  These dust characteristics are appropriate for the dense and cold environment of massive star formation.

\footnotetext[1]{The opacity table for the $10^5$ cm$^{-3}$ molecular density can be found at http://hera.ph1.uni-koeln.de/$\sim$ossk/Jena/dust.html}

We derived best-fit parameters --- $T_{\rm d}$ and $N_{\rm H_2}$ --- using least-squares fitting. An error of 10\% is assumed for the flux measured at each band. Approximately 12\% of the sources were not covered by the PACS instrument at 160 $\mu$m. For these sources, only the SPIRE data are used.

The uncertainties of the parameters were derived using the limits of the iso-contours of the chi-square function, as described in \citet{Lampton1976}.  The typical 1-$\sigma$ confidence interval is $[-10\%,+15\%]$ for the temperature, and $\pm25 \% $ for the column density for sources with data in the four bands considered. If only SPIRE was available, the fits become less reliable, with associated confidence intervals of $[-20\%,+30\%]$ and $\pm40\%$ for the dust temperature and column density, respectively.  

\begin{figure*}
%\rotate
\begin{center}
\epsscale{0.7}
\plotone{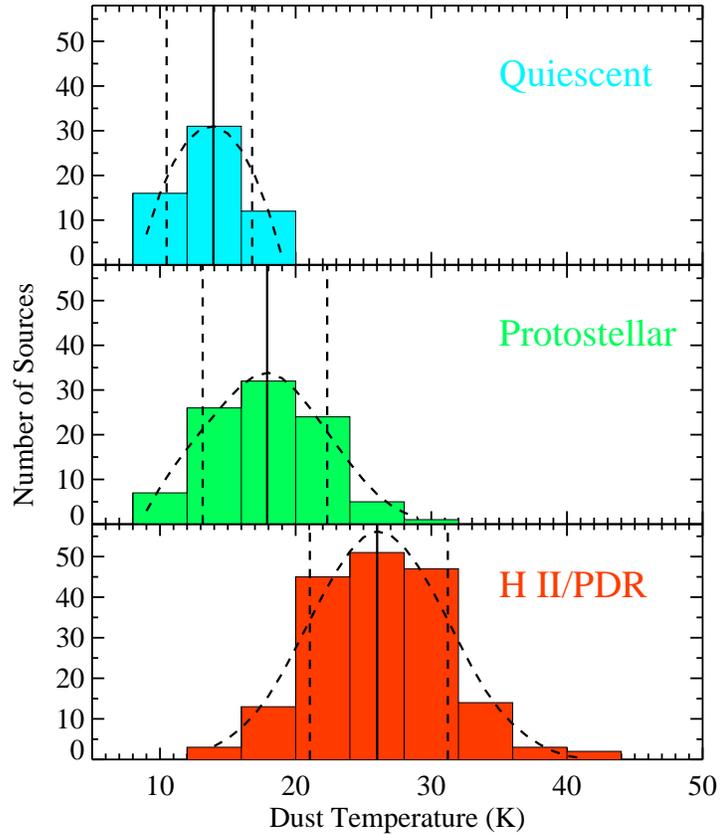}
\caption{Dust temperature distributions derived from {\it Herschel}/Hi-GAL data separated by evolutionary classification.  The median temperature for each stage is indicated by the solid vertical black line.  The Gaussian fits to the distributions along with the 1-$\sigma$ interval are overlayed as dashed lines as an estimate of the spread of the distributions. The median dust temperature increases as a function of the stage of evolution, as expected, indicating that the {\it Spitzer} IR classification scheme can be a reliable method of identifying the evolutionary stage of massive star formation.}
\label{temphist}
\end{center}
\end{figure*}

Dust temperatures and H$_2$ column densities were derived for 332 of the 333 sources.  We could not derive the parameters for one source that did not fall within the Hi-GAL survey range.  Figure \ref{temphist} shows the distribution of dust temperatures in each evolutionary category.  The solid vertical line indicates the median value of each distribution.  Gaussian curves were fit to the distribution of each classification to give a rough quantitative estimate of the spread and overlap of the distributions.  The curves and their 1-$\sigma$ intervals are indicated by dashed lines.

%Figure 2

For the temperature, a Kolmogorov-Smirnov (K-S) test, which determines whether two samples are drawn from the same parent population, shows that the distributions in the three evolutionary stages are drawn from distinct parent populations with a confidence higher than 99.9\%.  The gaussian fits to the distributions show that although there is overlap between the distributions, the median temperature of each category increases as a function of evolutionary stage, and the quiescent and \hii/PDR distributions do not overlap in the 1-$\sigma$ interval regions.

\begin{figure*}
%\begin{center}
%\epsscale{0.7}
%\plotone{figures/coldhist.eps}
\mbox{\subfigure{\includegraphics[width=3.5in]{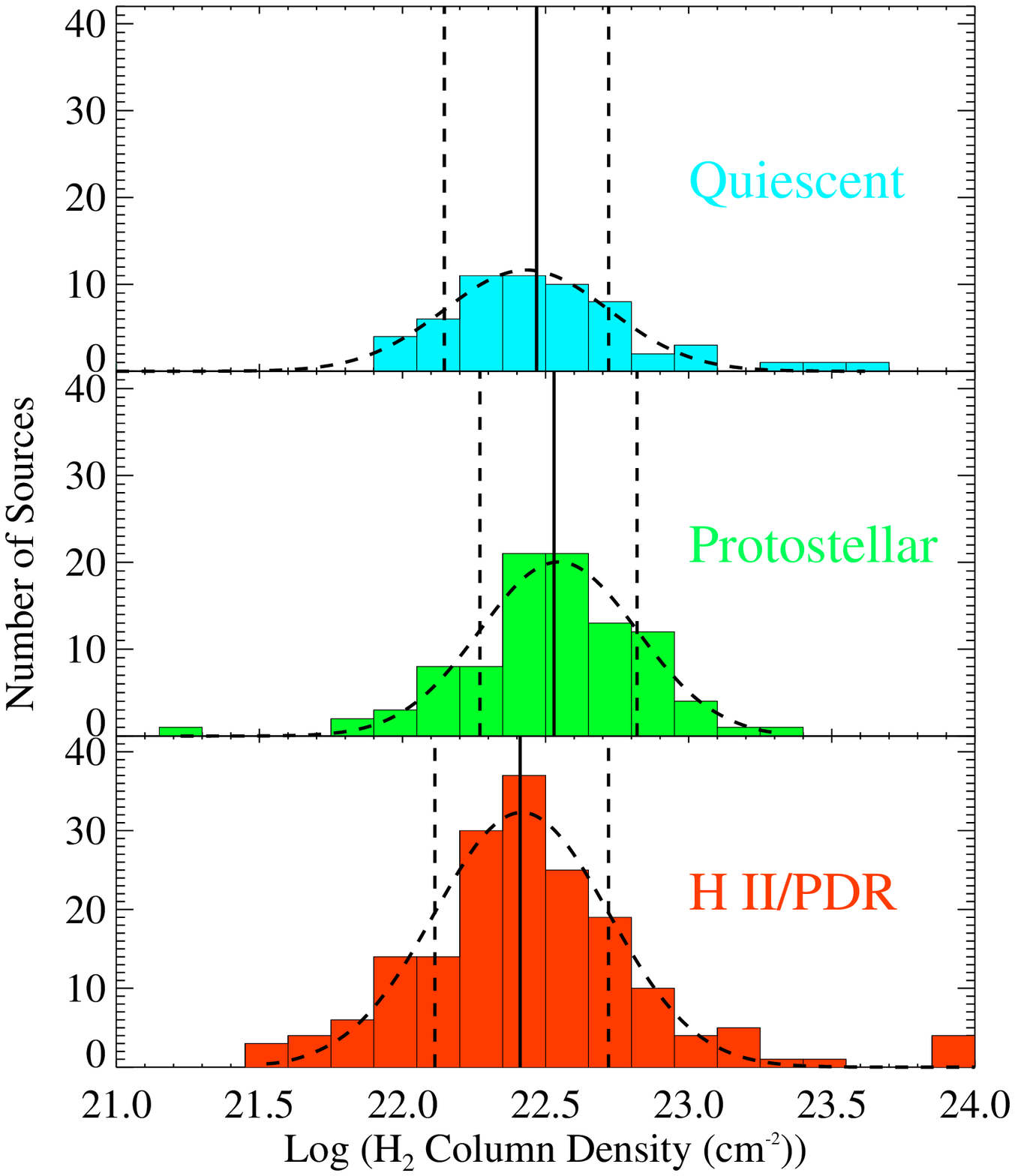}}}
%\epsscale{0.7}
%\plotone{figures/mphist.eps}
\subfigure{\includegraphics[width=3.5in]{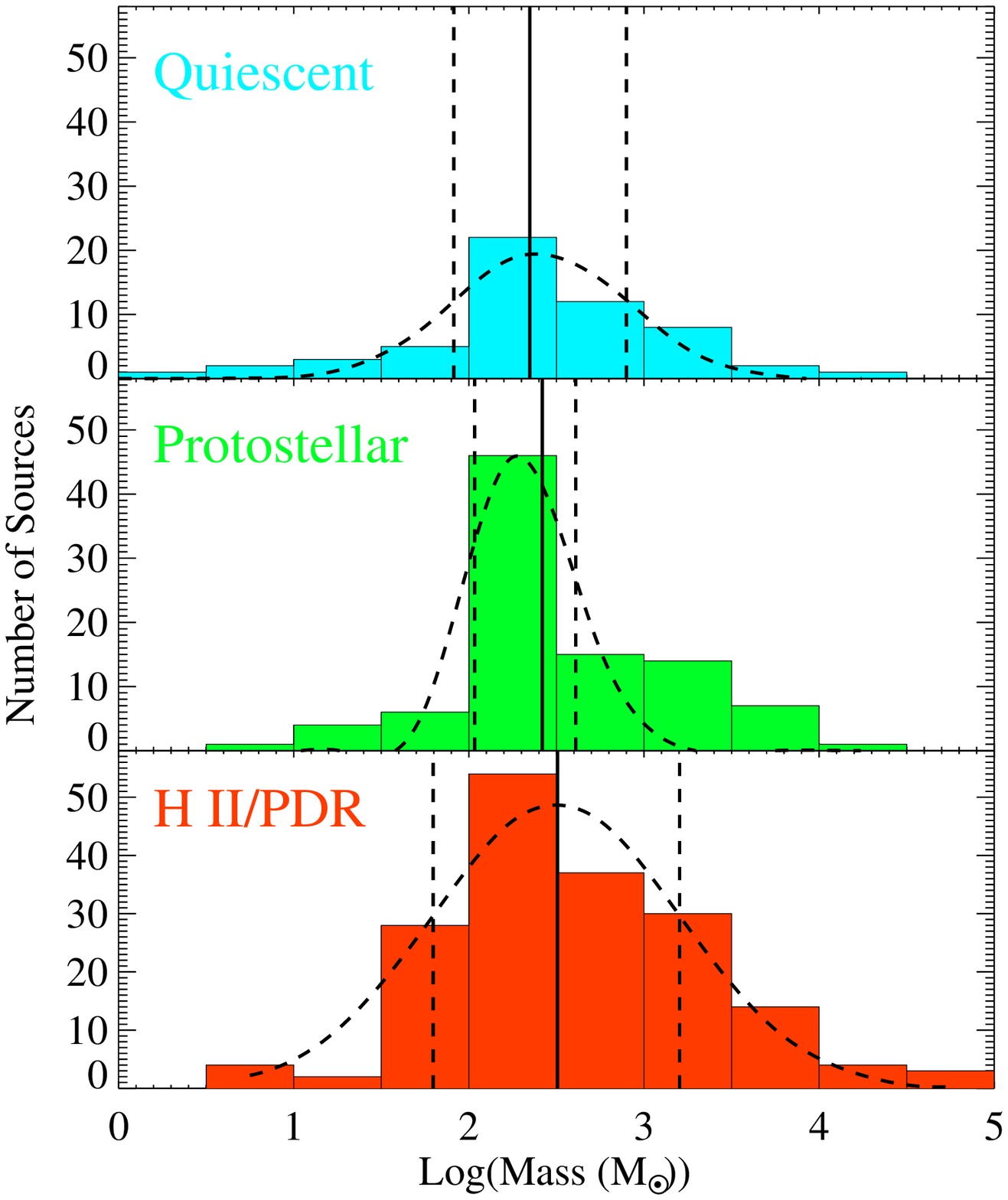}}
\caption{{\it Left}: H$_2$ column densities derived from {\it Herschel}/Hi-GAL dust emission separated by evolutionary stage.  Two sources are not plotted as their column densities are outside the plot range (one lower in the Quiescent distribution, one higher in the \hii/PDR distribution).  Medians values are indicated by the solid black vertical lines with Gaussian fits and 1-$\sigma$ interval overlayed as dashed lines.  {\it Right}: Masses of clumps derived from {\it Herschel}/Hi-GAL data separated by evolutionary classification, where the vertical solid black line indicates the median value for each stage with Gaussian fits and 1-$\sigma$ interval overlayed as dashed lines.}
\label{coldhist}
%\end{center}
\end{figure*}

The H$_2$ column densities and preliminary mass estimates separated by evolutionary stage are shown in Figure \ref{coldhist}.  For the H$_2$ column density, there is no strong trend with evolution and the distributions highly overlap.  A mass for each source within an estimated physical area covered by the circular angular area with diameter equivalent to one Mopra beam (38'') was derived from the H$_2$ column densities using preliminary kinematic distance estimates (Whitaker et al. 2013, in prep).  These preliminary masses are mostly larger than 10$^2$ M$_{\odot}$ and confirm that MALT90 is probing high-mass star-forming clumps.

%Figure 3
 
The dust temperature and H$_2$ column density of each source is listed in Table \ref{continuumtablesample}.  The median temperature, H$_2$ column density, and  mass in each evolutionary stage are listed in Table \ref{summarytable}.

%\begin{center}
\begin{deluxetable*}{ccccccc}
\tabletypesize{\scriptsize}
%\rotate
\tablecolumns{7}
\tablecaption{Summary of Results by Classification}
%\rotate
%\tablewidth{0pt}
\tablehead{
 \colhead{Property} &
 \multicolumn{2}{c}{\underline{~~~~~~~~~Quiescent~~~~~~~~~~}} & \multicolumn{2}{c}{\underline{~~~~~~~~Protostellar~~~~~~~~}} & \multicolumn{2}{c}{\underline{~~~~~~~~~~~~~~\hii/PDR~~~~~~~~~~~~~~}}\\
 \colhead{            } &
 \colhead{NS\tablenotemark{a}} &
 \colhead{Median} &
 \colhead{NS} &
 \colhead{Median} &
 \colhead{NS} &
 \colhead{Median}
}
\startdata
\nthp\ II Detections (K km s$^{-1}$) &  46  &  1.85  &  90  &  3.94  &  158  &  3.48 \\
       II Detections + Upper Limits (K km s$^{-1}$) & 51  & 1.72  & 94  & 3.76  & 173  & 3.10 \\
\tableline

\hcop\ II Detections (K km s$^{-1}$) &  43  &  1.66  &  87  &  2.59  &  164  &  3.76 \\
       II Detections + Upper Limits (K km s$^{-1}$) & 50  & 1.59  & 90  & 2.45  & 170  & 3.68 \\
\tableline

HCN II Detections (K km s$^{-1}$)  &  46  &  1.67  &  93  &  2.37  &  165  &  3.82 \\
    II Detections + Upper Limits (K km s$^{-1}$) & 51  & 1.42  & 94  & 2.37  & 173  & 3.59\\
\tableline
HNC II Detections (K km s$^{-1}$)  &  49  &  1.50  &  93  &  2.38  &  171  &  2.81 \\
    II Detections + Upper Limits (K km s$^{-1}$) & 52 & 1.37 & 95 & 2.37 & 177 & 2.67 \\
\tableline
\nthp/\hcop\ II Ratio---Detections  &  38  &  1.08  &  83  &  1.27  &  148  &  0.89 \\
%Detections + Upper Limits & 41 & 0.95 & 86 & 1.27 & 161 & 0.83\\
  Detections + Limits & 45 & 1.08 & 88 & 1.27 & 165 & 0.85\\
\tableline
HCN/HNC II Ratio---Detections  &  45  &  1.07  &  92  &  1.19  &  162  &  1.64 \\
%Detections + Upper Limits & 47 & 1.06 & 92 & 1.19 & 166 & 1.64\\
Detections + Limits & 47 & 1.06 & 93 & 1.19 & 168 & 1.64 \\
\tableline
\nthp\ Optical Depth---Detections  &  21  &  0.7  &  43  &  0.4  &  65  &  0.5\\
\tableline
\hcop\ Optical Depth---Detections  &  19  &  23.8  &  50  &  24.4  &  83  &  12.5 \\
%Detections + Lower Limits  &  24  &  35.73  &  66  &  24.62  &  107  &  15.3 \\
Detections + Limits  &  31  &  36.8  &  70  &  24.9  &  141  &  19.9 \\
\tableline
\nthp\ Column Density (cm$^{-2}$)---Detections  &  31  &  8.8 $\times$ 10$^{12}$  &  77  &  1.9 $\times$ 10$^{13}$  &  130  &  2.2 $\times$ 10$^{13}$ \\
Detections + Upper Limits &  51  &  6.1 $\times$ 10$^{12}$  &  94  &  1.5 $\times$ 10$^{13}$  &  172  &  1.7 $\times$ 10$^{13}$ \\
\tableline
\hcop\ Column Density (cm$^{-2}$)---Detections  &  19  &  1.3 $\times$ 10$^{14}$  &  50  &  2.5 $\times$ 10$^{14}$  &  83  &  2.4 $\times$  10$^{14}$ \\
%Detections + Lower Limits  &  24  & 1.5  $\times$ 10$^{14}$  &  66  &  2.6 $\times$ 10$^{14}$  & 107  &  2.7 $\times$  10$^{14}$ \\
Detections + Limits  &  31  &  1.5  $\times$ 10$^{14}$  &  70  &  2.7 $\times$ 10$^{14}$  &  141  &  3.1 $\times$ 10$^{14}$ \\
\tableline
\nthp\ Abundance---Detections  &  31  &  3.0 $\times$ 10$^{-10}$  &  77  &  5.0 $\times$ 10$^{-10}$  &  130  &  7.7 $\times$ 10$^{-10}$ \\
Detections + Upper Limits  &  51  &  2.6 $\times$ 10$^{-10}$  &  94  &  4.9 $\times$ 10$^{-10}$  &  172  &  7.6 $\times$ 10$^{-10}$ \\
\tableline
\hcop\ Abundance---Detections  &  19  &  4.0 $\times$ 10$^{-9}$  &  50  &  6.4 $\times$ 10$^{-9}$  &  83  &  7.5 $\times$ 10$^{-9}$ \\
%Detections + Lower Limits  &  24  &  5.0 $\times$ 10$^{-9}$  &  66  &  6.6 $\times$ 10$^{-9}$  &  107  &  8.1 $\times$ 10$^{-9}$ \\
Detections + Limits  &  31  &  5.3 $\times$ 10$^{-9}$  &  70  &  6.6 $\times$ 10$^{-9}$  &  141  &  9.9 $\times$ 10$^{-9}$ \\
\tableline
\nthp/\hcop Abundance Ratio---Detections  &  15  &  0.07  &  47  &  0.08  &  76  &  0.11 \\
 %Detections + Lower Limits &  19  &  0.07  &  48  &  0.08  &  94  &  0.09 \\
 Detections + Limits &  28  &  0.04  &  67  &  0.07  &  122 &  0.08 \\
\tableline
Dust Temperature (K)  &  59  &  13.9  &  95  &  17.9  &  178  &  26.0 \\
Clump Mass (M$_{\odot}$)  &  57  &  220  &  94  &  260  &  176  &  320 \\
H$_2$ Column Density (cm$^{-2}$)  &  59  &  2.9 $\times$ 10$^{22}$  &  95  &  3.4 $\times$ 10$^{22}$  &  178  &  2.6 $\times$ 10$^{22}$ \\
\enddata
\label{summarytable}
\tablenotetext{a}{NS = Number of Sources}
\tablecomments{II = Integrated Intensity}
\end{deluxetable*}
%\end{center}

\subsection{Determination of Molecular Abundances}

\subsubsection{Determining Detections of Molecular Lines}
\label{determining detections}
Table \ref{detectiontable} summarizes the number of detections and nondetections for the \nthp, \hcop, and \htcop\ (1-0) lines.  To determine which sources had significant detections of each line, Gaussian curves were fit to the average spectrum inside the central area of diameter 38'' of each line map.  We disregarded sources (15 for \nthp, 23 for \hcop) where the lines were either too broadened (mostly galactic center sources, line width ($\Delta$v) $\gtrsim$10 km s$^{-1}$), located too close to the edge of the spectral window to give reliable fits, or showed multiple velocity components that could not be fit.  These sources are not include in further analysis.

\begin{center}
\begin{deluxetable*}{cccc}
\tabletypesize{\scriptsize}
\tablecolumns{4}
\tablecaption{Detection Statistics}
%\rotate
\tablewidth{0pt}
\tablehead{
 \colhead{Molecular Line} &
 \colhead{Detections} &
 \colhead{Non Detections} &
 \colhead{Sources Not Used}
}
\startdata
\nthpone\  &  238  &  80  &  15  \\
\hcopone\  &  284  &  26  &  23  \\
\htcop\ (1-0) &  214  &  96  &  5\tablenotemark{a}  \\
HNC (1-0)\tablenotemark{b}  &  313  &  11  &  9   \\
HCN (1-0)\tablenotemark{b}  &  304  &  14  &  15\tablenotemark{c}  \\ 
\enddata
\tablenotetext{a}{The five sources where \htcop\ (1-0) was flagged as problematic were also flagged in \hcopone.}
\tablenotetext{b}{The HNC and HCN (1-0) detections are of their integrated intensity values, as described in Section \ref{intensity ratio results}.}
\tablenotetext{c}{The sources that were not used for HCN are the same sources for \nthp\ because the line widths used to determine the HCN (1-0) integrated intensities are those of the \nthp\ 123-012 HF (see Section \ref{intensity ratio results}).}
\label{detectiontable}
\end{deluxetable*}
\end{center}

We measured the integrated intensity (I = $\int T_{A} dv$) of each line within $\pm$1 km s$^{-1}$ around the V$_{lsr}$ determined by the Gaussian fits.  The 1-$\sigma$ uncertainty of this integrated intensity, $\sigma_{D}$, was determined from the T$_{\rm rms}$ of the antenna temperature measured from the signal-free part of the spectrum.  For each source, we consider that a line has been detected if the integrated intensity within the $\pm$1 km$s^{-1}$ interval is greater than 3$\sigma_{D}$.

\subsubsection{Integrated Intensities}
\label{intensity ratio results}
After determining if a line was detected, we measured the integrated intensities of \nthpone, \hcopone, \hcnone, and \hncone\ over a larger velocity range.  For \hcop\ and HNC, the line intensities were integrated over the range V$_{lsr}$ $\pm$ $\Delta$$v$ of the Gaussian fit.  For sources where the line was not detected, a line width of 2.0 km s$^{-1}$ was assumed and a limit to the interated intensity was determined by integrating over the range V$_{lsr}$ $\pm$ 2 km s$^{-1}$.  The HNC integrated intensity was not determined for nine sources as their line widths were either larger than $\sim$10 km s$^{-1}$ or the spectra showed multiple velocity components that could not be reliably fit.  For the two molecules that exhibit hyperfine splitting, \nthp\ and HCN, we measured the integrated intensity over the range:
\begin{equation}
 v_{1{\rm HF}} -\Delta~v_{{\rm N_{2}H^+}_{123-012}}~{\rm to}~v_{2{\rm HF}} +\Delta~v_{{\rm N_2H^+}_{123-012}}, 
\end{equation}
where $\Delta$$v$$_{{\rm N_2H^+}_{123-012}}$ is the line width of the \nthp\ 123-012 hyperfine (HF) transition and $v$$_{1{\rm HF}}$ and $v$$_{2{\rm HF}}$ are defined as the velocities of the outer hyperfine lines of each molecule.  For sources where the \nthp\ 123-012 HF line was not detected, a line width of 2.0 km s$^{-1}$ was assumed and upper limits of the integrated intensities were found for \nthp\ and HCN.  The integrated intensity detections of HCN and HNC are listed in Table \ref{detectiontable}.  The  integrated intensity ratios of \nthpone\ to \hcopone\ [I(\nthp)/I(\hcop)] and of \hcnone\ to \hncone\ [I(HCN)/I(HNC)] were then determined.

Of the 333 sources, we did not derive  I(\nthp)/I(\hcop) for 35 sources and I(HCN)/I(HNC) for 25 sources as they were either deemed problematic, as stated in the previous section, or the integrated intensities of both lines in the ratio were upper limits.  For both integrated intensity ratios, the number of sources in each evolutionary stage and their median values are listed in Table \ref{summarytable}.  The typical uncertainties of the ratios range from $\sim$7--12\% for I(\nthp)/I(\hcop) and $\sim$6--13\% for I(HCN)/I(HNC).

\begin{figure*}
%\begin{center}
%\epsscale{0.7}
%\plotone{figures/ration2hphcophist.eps}
\mbox{\subfigure{\includegraphics[width=3.5in]{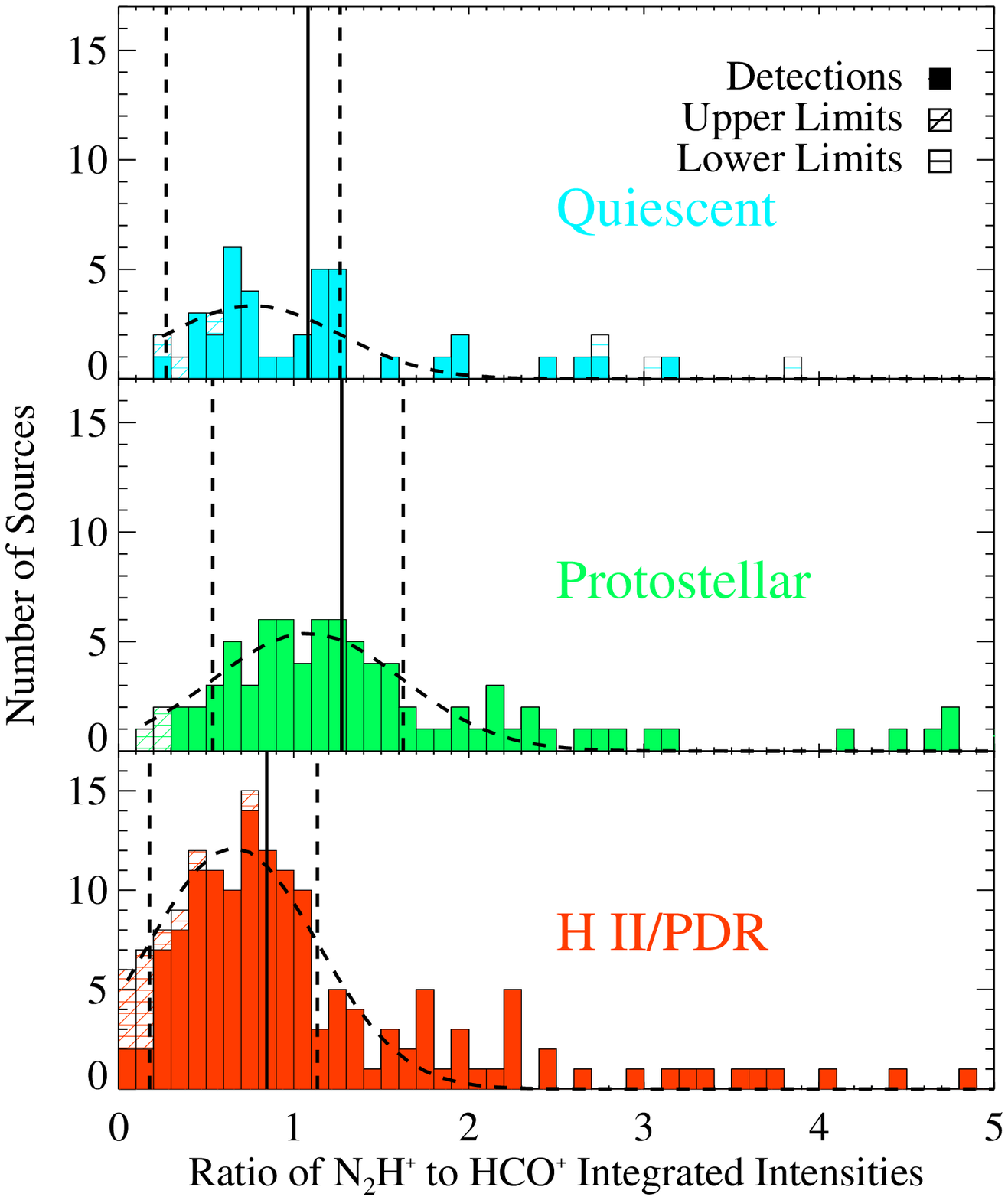}}}
%\epsscale{0.7}
%\plotone{figures/ratiohcnhnchist.eps}
\subfigure{\includegraphics[width=3.5in]{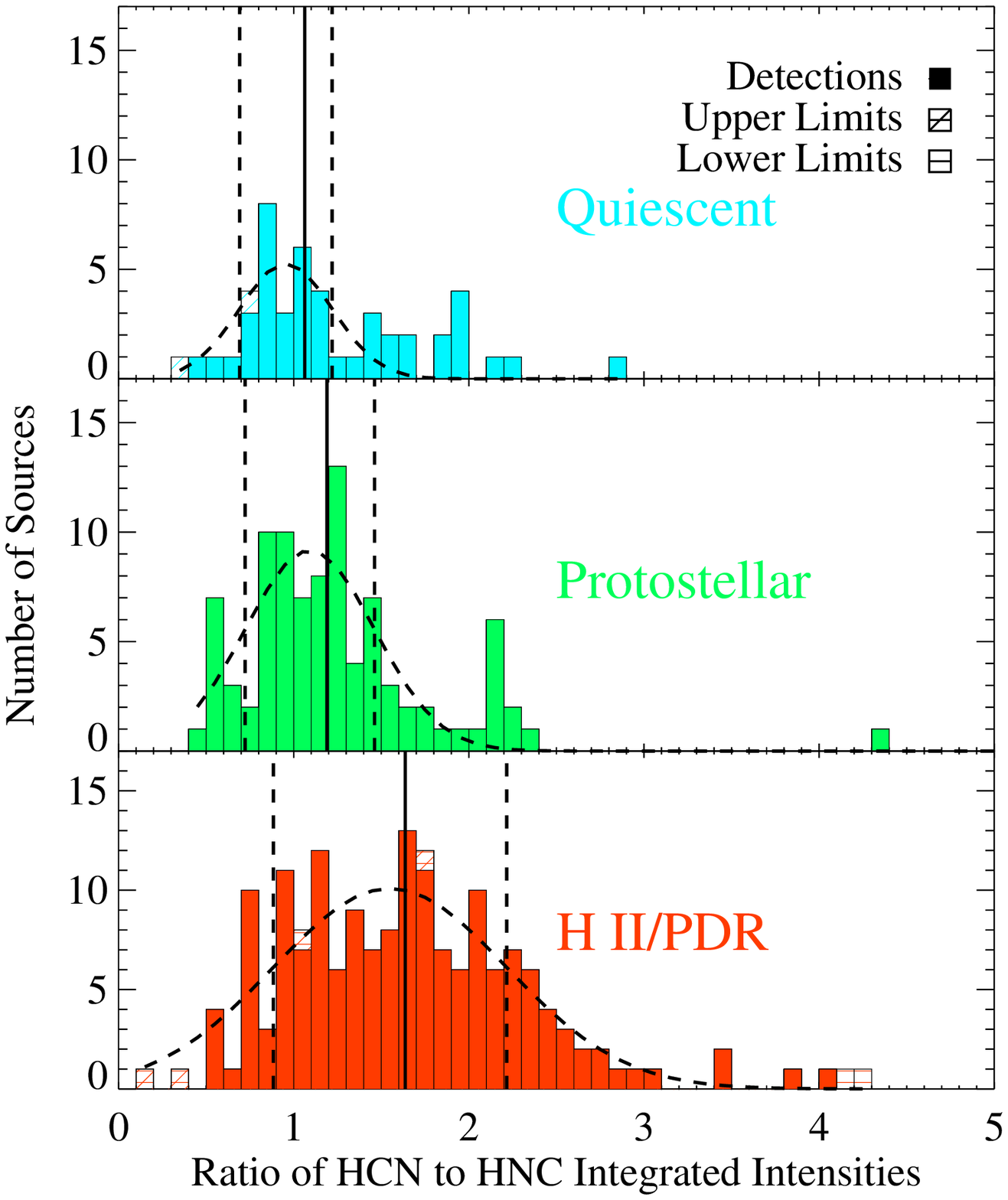}}
\caption{{\it Left}: \nthp\ to \hcop\ integrated intensity ratio distributions separated by classification, which show no clear trend with evolution.  The ratios of 13 sources are not displayed in the plot because they are larger than 5.  {\it Right}: HCN to HNC integrated intensity ratios separated by classification.  The median integrated intensity ratios may suggest an increase between the earliest and latest evolutionary stages, but the three distributions heavily overlap. The ratio of 1 source is not displayed in the plot because it is larger than 5.  Solid black vertical lines show the median values with Gaussian fits and 1-$\sigma$ interval overlayed as dashed lines. }
\label{ration2hphcophcnhnchist}
%\end{center}
\end{figure*}

Figure \ref{ration2hphcophcnhnchist} shows the values of I(\nthp)/I(\hcop) and I(HCN)/I(HNC) for the MALT90 clumps separated into their different evolutionary classifications.  The median values of the distributions, including detections, upper and lower limits, are indicated by black solid lines.  Gaussian curves fit to the distributions and their 1-$\sigma$ intervals are indicated by dashed lines.  For I(\nthp)/I(\hcop), the median values for detections of the quiescent, protostellar, and \hii/PDR distributions are 1.1, 1.3, and 0.9, respectively.  About 18\% of the MALT90 sources were visually determined to show features of self-absorption in the \hcop\ spectrum.  The median value of sources where the \hcop\ spectrum was self-absorbed was slightly higher than for sources where \hcop\ was not self-absorbed.  The 1-$\sigma$ intervals of the Gaussian fits to the three distributions of I(\nthp)/I(\hcop) overlap and do not show any clear trend as a function of evolutionary stage.

%Figure 4

The median values of the ratio I(HCN)/I(HNC) for detections for quiescent, protostellar, and \hii/PDR sources are 1.1, 1.2, and 1.6, respectively.  Similar to the I(\nthp)/I(\hcop), the 1-$\sigma$ intervals of the Gaussian fits overlap significantly in the three distributions.

The number of sources in each evolutionary stage and their median values are listed in Table \ref{summarytable} for both integrated intensity ratios.

%For \nthp, 232 sources were detections, 88 sources could not be detected above 3$\sigma_{II}$.  
\subsubsection{Molecular Optical Depths}
\label{optical depth}

The optical depths for \nthp\ and \hcop\  were calculated by assuming local thermodynamic equilibrium (LTE), and by assuming that the excitation temperatures of the lines were equal to the derived dust temperature.  We calculated the optical depths of \nthp\ and \hcop\ following the procedure described in \citet{Sanhueza2012}.  In the case of \nthp, we found the opacity by taking the ratio of the brightness temperatures of its JF$_1$F = 123-012 and JF$_1$F = 112-012 HF transitions.  For \hcop, we used its optically thin isotopologue, \htcop, whose ground state transition was also observed as part of MALT90.  Based on whether the \nthpone, \hcopone, and \htcop\ (1-0) lines were detected, for some sources, we only derived limits on the optical depths of \nthp\ and \hcop, and for some other sources, we did not derive optical depths at all, as described below.  

For \nthp, we did not calculate an optical depth for sources where both HFs were not detected or where T$_{123-012}$ $\ge$ $\frac{5}{3}$T$_{112-012}$, where T$_{123-012}$ and T$_{112-012}$ are the \nthp\ 123-012 and 112-012 HF line temperatures measured at the V$_{lsr}$.  The former group of sources are treated as upper limits when calculating column densities and abundances.  The latter group is composed of significant detections of the line, but the ratio of the HF line temperatures is larger than 5/3 within the uncertainties.  As 5/3 is the ratio of the statistical weights of the two HF lines \citep{Daniel2006}, line temperature ratios larger than 5/3 between the two HF lines are unphysical.  For sources in both groups above, we assume that the line is optically thin.  For \hcop, optical depths were not calculated if the \hcopone\ line was not detected or if T$_{{\rm H^{13}CO^{+} (1-0)}}$ $\ge$ T$_{{\rm HCO^{+} (1-0)}}$.  In the latter case, the presense of \htcop\ indicates that \hcop\ is likely present but significantly self-absorbed.  Upper limits to the optical depth were derived for sources where \hcopone\ was detected but \htcop\ (1-0) was not.  The median optical depth values for detections and limits for \nthp\ and \hcop\ are listed in Table \ref{summarytable}.

For 22\% of the sources where \hcop\ was self-absorbed ($\sim$4\% of the total 333 sources), T$_{{\rm H^{13}CO^{+} (1-0)}} \ge$ T$_{{\rm HCO^{+} (1-0)}}$, and optical depths were not derived.  For the other 78\% ($\sim$14\% of total 333 sources), the derived optical depths are treated as lower limits.  We did not derive \hcop\ optical depths for six sources where \hcop\ was self-absorbed, thus a lower limit, and \htcop\ was not detected, which would be an upper limit.

The line fit parameters for the \nthp\ 123-012 and 112-012 HFs, \hcop, and \htcop\ lines for the 333 sources can be found in the online material.  The V$_{lsr}$ and $\Delta$$v$ values listed were found by fitting the average spectra within the central area of diameter 38''.  The consensus velocities of the MALT90 sample will be released in a forthcoming publication (J. S. Whitaker et al., in preparation).  These velocities may vary from the ones listed in the present study by $\sim$1 km s$^{-1}$.  A sample of the \nthp\ 123-012 HF line fit parameters is shown in Table \ref{n2hpmiddlelinetablesample}.

\begin{center}
\begin{deluxetable*}{cccccccc}
\tabletypesize{\scriptsize}
\tablecolumns{8}
\tablecaption{Gaussian Fit Parameters for \nthp\ 123-012 HF Line\label{n2hpmiddlelinetablesample}}
%\rotate
\tablewidth{0pt}
\tablehead{
 \colhead{Source} &
 \colhead{T$_{rms}$} &
 \colhead{T$^{*}_A$} &
 \colhead{$\sigma_{D}$} &
 \colhead{II over} &
 \colhead{$\Delta$V} &
 \colhead{V$_{lsr}$\tablenotemark{a}} &
 \colhead{Detection}\\
 \colhead{      } &
 \colhead{(K)} &
 \colhead{(K)} &
 \colhead{(K km s$^{-1}$)} &
 \colhead{$\pm$1 km s$^{-1}$ of T$^{*}_A$} &
 \colhead{(km s$^{-1}$)} &
 \colhead{(km s$^{-1}$)} &
 \colhead{Category}
}
\startdata
G000.253+00.016   &   $\cdots$   &   $\cdots$   &   $\cdots$   &   $\cdots$     &   $\cdots$     &   $\cdots$   &   Broadened\tablenotemark{b}   \\
G000.414+00.045   &   $\cdots$   &   $\cdots$   &   $\cdots$   &   $\cdots$     &   $\cdots$     &   $\cdots$   &   Broadened   \\
G003.033+00.405   &   0.2   &   $\cdots$   &   0.1   &   0.08     &   2.00     &   12.2$\pm$0.9   &   Upper Limit\tablenotemark{c}  \\
G003.089+00.164   &   $\cdots$   &   $\cdots$   &   $\cdots$   &   $\cdots$     &   $\cdots$     &   $\cdots$   &   Broadened   \\
\enddata
\tablenotetext{a}{These v$_{lsr}$ values are based on the fits to the average spectra in the central circular area of diameter 38'' of each map.  The consensus velocities of the MALT90 sample will be released in a forthcoming publication (Whitaker et al. 2013, in prep), which may differ from the present v$_{lsr}$ values by $\sim$1 kms$^{-1}$.}
\tablenotetext{b}{For sources categorized as broadened, their lines were too broad to produce reliable fits and so these sources are not included in the analysis of the molecular parameters.  These sources are mostly located near the galactic center.}
\tablenotetext{c}{Sources labeled as Upper Limits were non detections, but a $\Delta$V of 2.0 km s$^{-1}$ was assumed and the brightness temperatures of the lines were set equal to 3 times the rms of the spectrum to produce upper limits.  Therefore, no uncertainties are listed for these values.}
\tablecomments{This table is available in its entirety in a machine-readable form in the online journal. A portion is shown here for guidance regarding its form and content.}
\end{deluxetable*}
\end{center}

\subsubsection{Column Densities and Abundances}
\label{column densities and abundances}
As with the optical depths, the column densities for \nthp\ and \hcop\ were derived following the procedure described in \citet{Sanhueza2012}.  We did not derive column densities or abundances for \hcop\ if it was not detected.  For sources where the \hcopone\ line was detected, but the \htcop\ (1-0) line was not, upper limits were computed by setting the brightness temperature of the \htcop\ (1-0) line equal to 3T$_{rms}$.

\begin{figure*}
%\begin{center}
%\epsscale{0.7}
%\plotone{figures/abundancen2hphist.eps}
\mbox{\subfigure{\includegraphics[width=3.5in]{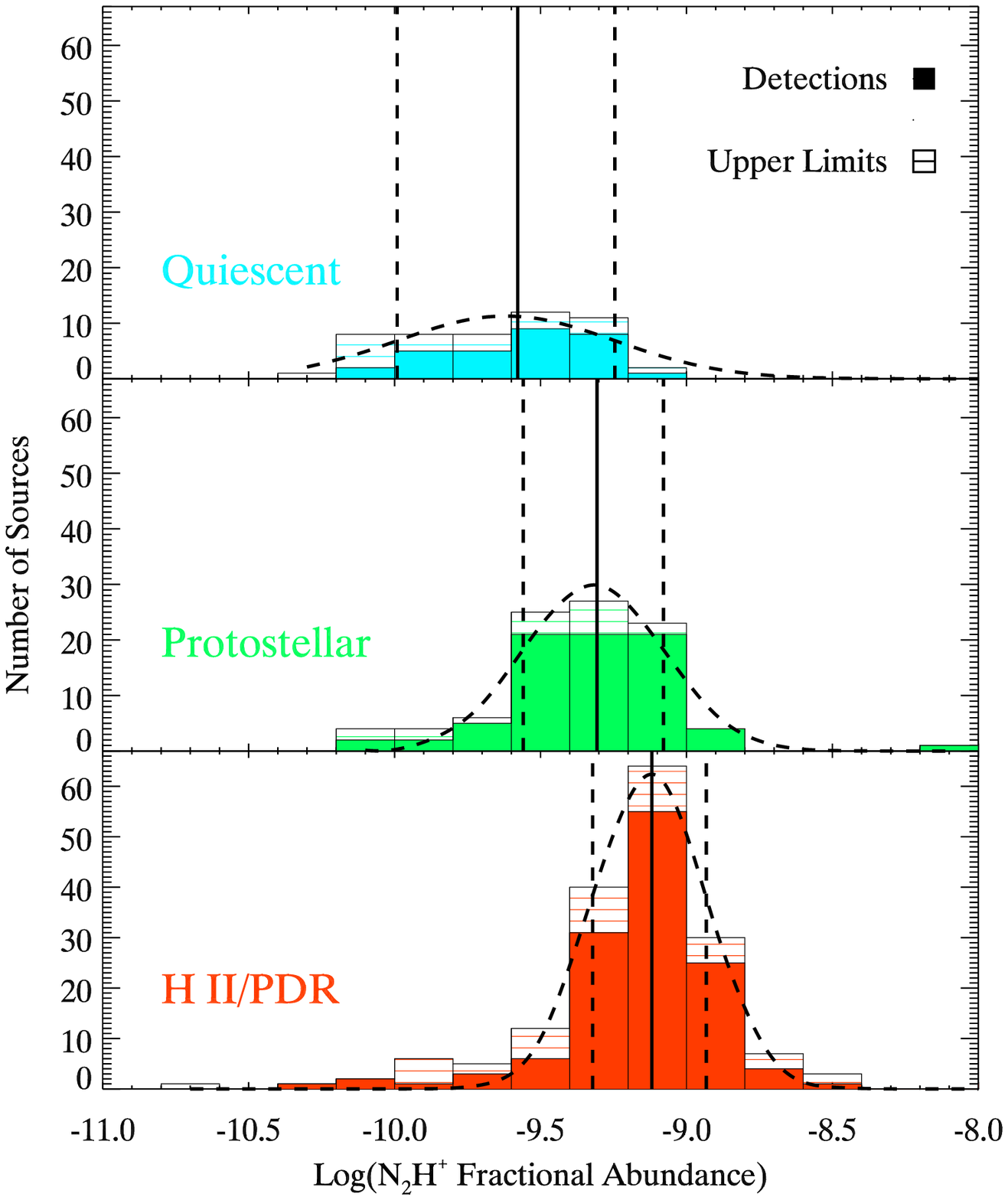}}}
%\epsscale{0.7}
%\plotone{figures/abundancehcophist.eps}
\subfigure{\includegraphics[width=3.5in]{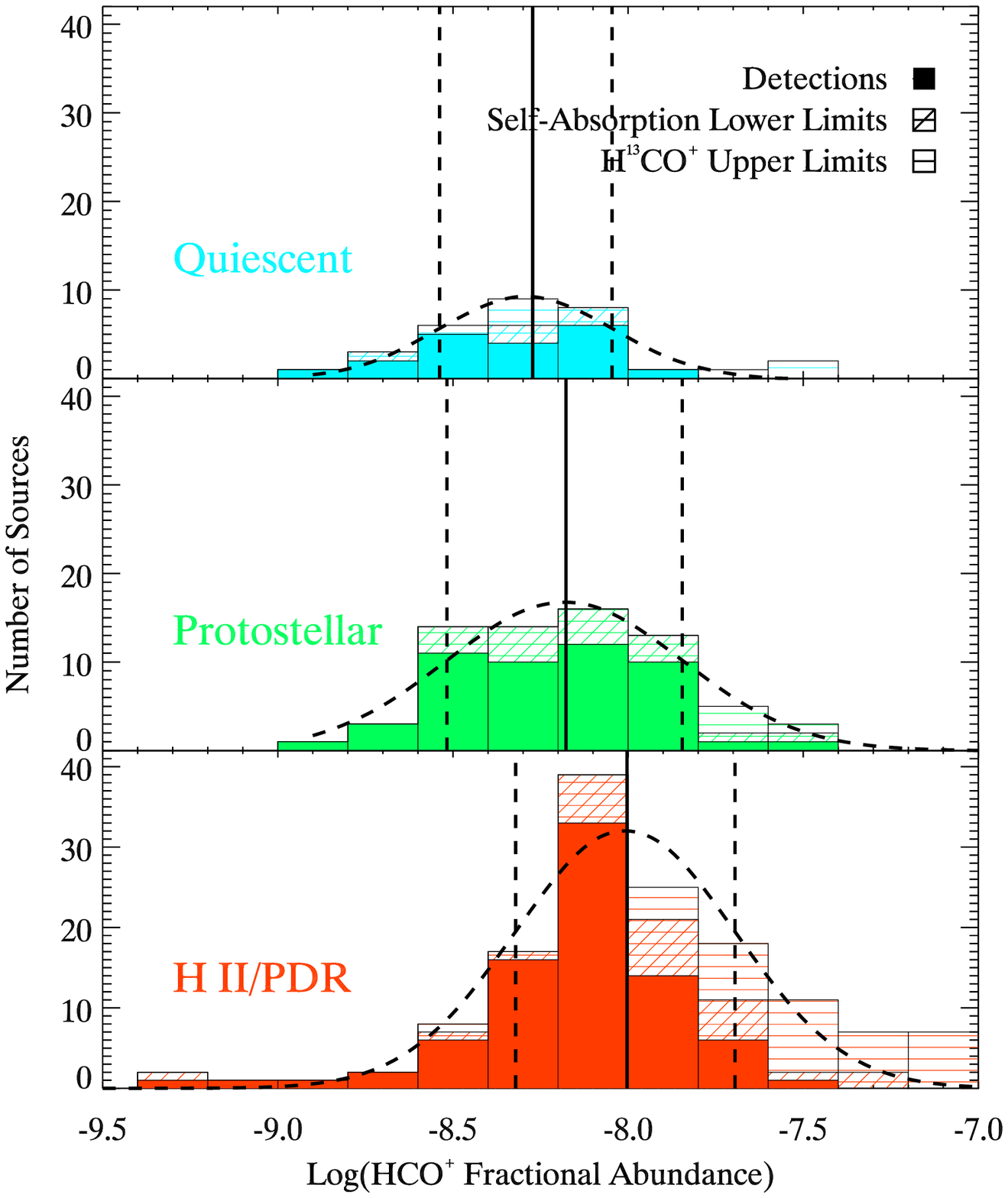}}
\caption{{\it Left}: \nthp\ abundances separated by classification.  Upper limits are sources where either the \nthp\ 123-012 HF or the 112-012 HF line was not detected and column densities and abundances were determined by assuming the sources were optically thin.  Overall, the \nthp\ abundance increases as a function of evolution, although there is some overlap between the distributions.  Two sources are not included in the plot as their abundances are larger than the plot range.  {\it Right}: \hcop\ abundances separated by classification.  Lower limits are sources where the \hcopone\ spectra showed self-absorbed profiles, and upper limits are sources where the \hcopone\ line was a detection but \htcop\ (1-0) was not.  Similar to the \nthp\ abundance distributions, the \hcop\ abundances increase as a function of evolution, but with overlap between the distributions.  Four sources are not included in the plot as their abundances are larger than the plot range.}
\label{n2hphcopabundancehist}
%\end{center}
\end{figure*}

The \nthp\ and \hcop\ abundances were found by dividing the \nthp\ and \hcop\ column densities by the H$_2$ column density determined from {\it Herschel} Hi-GAL dust continuum emission (see Section \ref{dust temperatures and h2 column densities}).  The median values of the \nthp\ and \hcop\ column densities and abundances are listed in Table \ref{summarytable}.  Figure \ref{n2hphcopabundancehist} shows the distributions of both \nthp\ and \hcop\ abundances separated by their {\it Spitzer} IR classification, where, similar to Figure \ref{temphist}, gaussian curves are fit to the distributions, and the fit and 1-$\sigma$ intervals are indicated by dashed lines.  The median values of the distributions are indicated by the solid vertical line.

%figure 5

If the lower limits of the abundances of the self-absorbed sources are added to the detections for \hcop, the median value of the quiescent sources increases by about 24\%.  The protostellar distribution remains mostly unaffected, and the \hii/PDR distribution median value increases by nearly 7\%.

The MALT90 \nthp\ and \hcop\ abundances are in agreement with studies of massive clumps in IRDCs by \citet{Sakai2008}, \citet{Vasyunina2011}, and \citet{Sanhueza2012}.  \citet{Sanhueza2012}, who categorized their sample into different evolutionary stages, found that both \nthp\ and \hcop\ abundances increase as a function of evolutionary stage, in agreement with our findings.  \citet{Liu2013} performed a similar study of a subsample of MALT90 clumps.  Their derived column densities are lower than the ones derived in this study, however, as they assumed a beam filling factor of one in their analysis.

The \nthp\ and \hcop\ molecular properties of each source are listed in Table \ref{derivedmoleculartablesample}.

\begin{center}
\begin{deluxetable*}{ccccccccc}
\tabletypesize{\scriptsize}
\tablecolumns{9}
\tablecaption{Derived Molecular Properties\label{derivedmoleculartablesample}}
%\rotate
\tablewidth{0pt}
\tablehead{
 \colhead{Source} &
 \multicolumn{2}{c}{\underline{~~~Optical Depth~~}} & \multicolumn{2}{c}{\underline{~~~~~Column Density~~~~~}} & \multicolumn{2}{c}{\underline{~~~~~Abundance~~~~~}} & \multicolumn{2}{c}{\underline{~~~~~~~~Detection\tablenotemark{a}~~~~~~~~}}\\
 \colhead{Name} &
 \colhead{\nthp} &
 \colhead{\hcop} &
 \colhead{\nthp} &
 \colhead{\hcop}&
 \colhead{\nthp} &
 \colhead{\hcop} &
 \colhead{\nthp} &
 \colhead{\hcop} \\
 \colhead{     }&
 \colhead{     }&
 \colhead{     }&
 \colhead{($\times$10$^{12}$ cm$^{-2}$)} &
 \colhead{($\times$10$^{14}$ cm$^{-2}$)} &
 \colhead{($\times$10$^{-10}$)} &
 \colhead{($\times$10$^{-9}$)} &
 \colhead{     } &
 \colhead{     }
}
\startdata
G000.253+00.016   &      $\cdots$     &  $\cdots$    &   $\cdots$   &   $\cdots$   &   $\cdots$   &   $\cdots$   &   Broadened   &   Broadened   \\
G000.414+00.045   &      $\cdots$     &  $\cdots$    &   $\cdots$   &   $\cdots$   &   $\cdots$   &   $\cdots$   &   Broadened   &   Broadened   \\
G003.033+00.405   &      $\cdots$     &  $\cdots$    &   7.4  &  $\cdots$   &   3.0  &  $\cdots$   &   Upper Limit   &   Broadened   \\
G003.089+00.164   &      $\cdots$     &  $\cdots$    &   $\cdots$   &   $\cdots$   &   $\cdots$   &   $\cdots$   &   Broadened   &   Broadened   \\
\enddata
\tablenotetext{a}{Detections refer to whether the \nthpone\ and \hcopone\ lines were detected as determined in Section \ref{determining detections}. For sources categorized as broadened, their lines were too broad to produce reliable fits and so these sources are not included in the analysis of the molecular parameters.  These sources are mostly located near the galactic center.  Sources labeled as Upper Limits were non detections, but a $\Delta$V of 2.0 km s$^{-1}$ was assumed and the brightness temperatures of the lines were set equal to 3 times the rms of the spectrum to produce upper limits.}
\tablecomments{This table is available in its entirety in a machine-readable form in the online journal. A portion is shown here for guidance regarding its form and content.}
\end{deluxetable*}
\end{center}

\section{Discussion}

\subsection{Dust Temperatures and H$_2$ Column Densities}
Figure \ref{temphist} shows that the dust temperatures increase as a function of evolution, although there is overlap between the different categories.  Some overlap is expected because the distributions are a binned representation of a continuous process.  The quiescent and protostellar distributions are more similar to each other than to the \hii/PDR distribution.  This is likely because the Mopra beam has an angular FWHM of 38'', and the point-like protostellar sources suffer greater beam dilution than the larger \hii\ regions.  Therefore, for protostellar sources, a larger amount of diffuse, likely cold material surrounding the central source contributes to the flux measured in the beam, which makes this sample more similar to the cold, quiescent group.

Overall, the dust temperatures increase as a function of evolutionary stage, which gives confidence to the {\it Spitzer} IR-based classification scheme as a reliable method of classifying the evolutionary stage of star-forming regions.

The distribution of H$_2$ column densities, shown in Figure \ref{coldhist} do not appear to have any correlation with evolutionary stage.  This unchanging distribution is consistent with the idea that we are probing the same types of objects are different stages of evolution.  Based on mass estimates derived from preliminary kinematic distances, the MALT90 sources are mostly high-mass clumps ($\sim$10$^2$--10$^3$ M$_{\odot}$), with typical masses of $\sim$200--300 M$_{\odot}$.

\subsection{Ratio of I(HCN) to I(HNC) as a Function of Evolution}
In cold clouds, the relative abundance ratio of HCN to HNC [X(HCN)/X(HNC)] is observed to be near unity \citep{Wootten1978, Hirota1998, Hickman2005, Padovani2011}, in agreement with model predictions \citep{Herbst2000, Sarrasin2010}.  In their study of 19 dark cloud cores, \citet{Hirota1998} found that X(HCN)/X(HNC) was generally lower than unity, and also found that the ratio increases as the temperature in a source rises.  This result was also found by \citet{Goldsmith1981, Goldsmith1986}, \citet{Irvine1984}, and \citet{Churchwell1984}.  \citet{Tennekes2006}, in their study of the protostellar core Chameleon-MMS1, found that the abundance ratio of HCN/HNC to be $\sim$0.35, and \citet{Schilke1992} found that X(HCN)/X(HNC) to be $\sim$100 in the Orion hot core region.  Examined together, these studies suggest that the HCN/HNC abundance ratio of a star-forming region increases as it evolves to higher temperatures.

We determined whether the same trend was found in the MALT90 clumps by comparing the I(HCN)/I(HNC) as a function of evolutionary stage.  We did not derive their abundance ratios because an optically thin isotopologue of HCN was not observed as part of MALT90 and the majority of the HCN spectra were too blended to perform hyperfine fitting.

The I(HCN)/I(HNC) measured for the MALT90 clumps, shown in Figure \ref{ration2hphcophcnhnchist}, suggest that the ratio increases as a function of evolution. The median values increase by $\sim$9\% and $\sim$33\% from the quiescent to the protostellar stage and from the protostellar to the \hii/PDR stage, respectively.  However, these difference are likely not significant, as there is substantial overlap between the distributions in each category.

\subsection{Comparison to Predictions of Chemical Evolution}
\label{comparison to predictions}
From previous studies and model predictions of chemical evolution in low-mass star-forming regions \citep[e.g.][]{Bergin1997, Lee2004}, we expected to find that the \nthp\ abundance of the MALT90 clumps would decrease as a function of evolutionary stage, while the \hcop\ abundance would increase.  These trends are based on the behavior of CO, which is expected to freeze out onto dust grains at the cold temperatures found in the early prestellar phases of star formation, and is released back into the gas phase as the central core warms up in the protostellar phase.  Since CO is a major destroyer of \nthp, the \nthp\ abundance should be relatively high when CO is trapped on dust mantles and should decrease as CO is released.  In the case of \hcop, as CO is the parent molecule, \hcop\ abundances should be depleted in the early, prestellar phase and should increase in the later, warmer phases when CO is released back into the gas phase.  

These trends are not what was observed for the clumps in MALT90.  Figure \ref{n2hphcopabundancehist} suggests that the \nthp\ abundance increases as a function of evolution, as was observed in a different sample by \citet{Sanhueza2012}.  Although the three distributions do significantly overlap, the median value of each clearly rises as a function of stage from quiescent to protostellar to \hii/PDR.  The \hcop\ abundance also rises as a function of evolution, which follows predictions.  

\begin{figure*}
\begin{center}
\epsscale{0.7}
\plotone{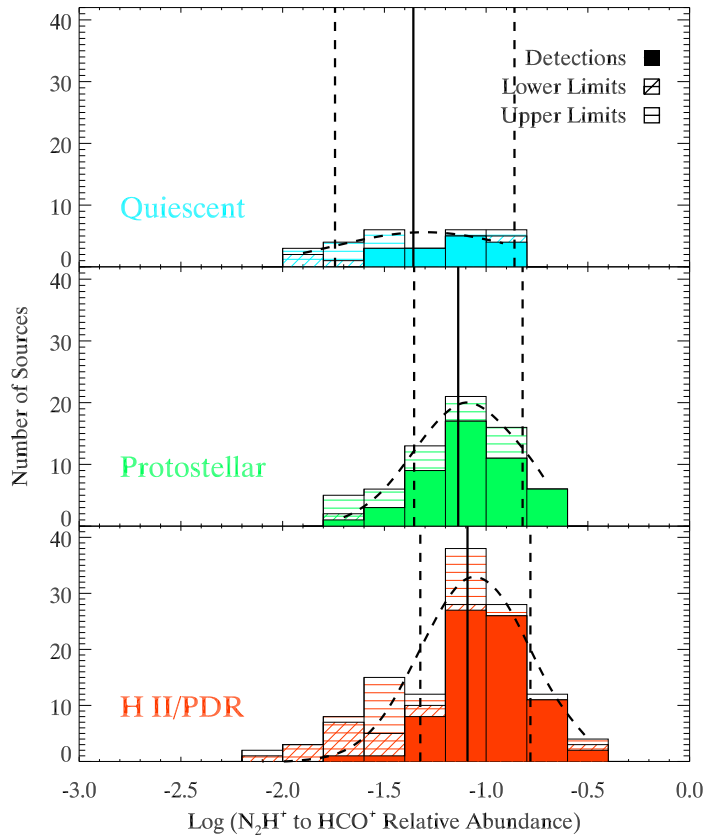}
\caption{Distribution of the ratios of \nthp\ to \hcop\ abundances separated by classification.  The median abundance ratios increase slightly as a function of evolution.  However, this trend is not statistically significant.}
\label{densityn2hphcophist}
\end{center}
\end{figure*}

We also predicted that the relative \nthp\ to \hcop\ abundance ratio would decrease as a function of evolutionary stage.  Our results, however, do not display this behavior (see Figure \ref{densityn2hphcophist}).  The \nthp/\hcop\ abundance ratio increases slightly as a function of evolution.  However, the 1-$\sigma$ intervals of the gaussian fits to the three distributions considerably overlap, suggesting that this trend is likely not significant.  

%figure 6

We identify two reasons why the results may be inconsistent with predictions.  First, the size scales probed by MALT90 are much larger than the scales probed by \citet{Bergin1997} and \citet{Lee2004}.  The Mopra beam is 38'', which at a distance of 3 kpc for high-mass star-forming regions, has a physical size of $\sim$0.6 pc.  Therefore, MALT90 probes star forming clumps, which may contain several star-forming cores (size $\le$ 0.1 pc), and also diffuse material.  In order to isolate only the material involved in star formation to compare to predictions of chemical evolution, we must probe individual cores inside clumps at higher angular resolution.

Second, chemical processes in high-mass star formation may differ from those of low-mass star formation. The physical conditions, such as temperature, density, and UV flux, involved in high-mass star forming regions are different than their low-mass counterparts.  However, again, to determine whether the inconsistency is due to a fundamental difference between the two regimes of star-formation, we need high angular resolution data.

\subsubsection{N$_{2}$H$^+$ Anomalies}
\label{anomalies}

\begin{figure*}
%\begin{center}
%\epsscale{0.5}
%\plotone{figures/Figure12.eps}
\mbox{\subfigure{\includegraphics[width=3.5in]{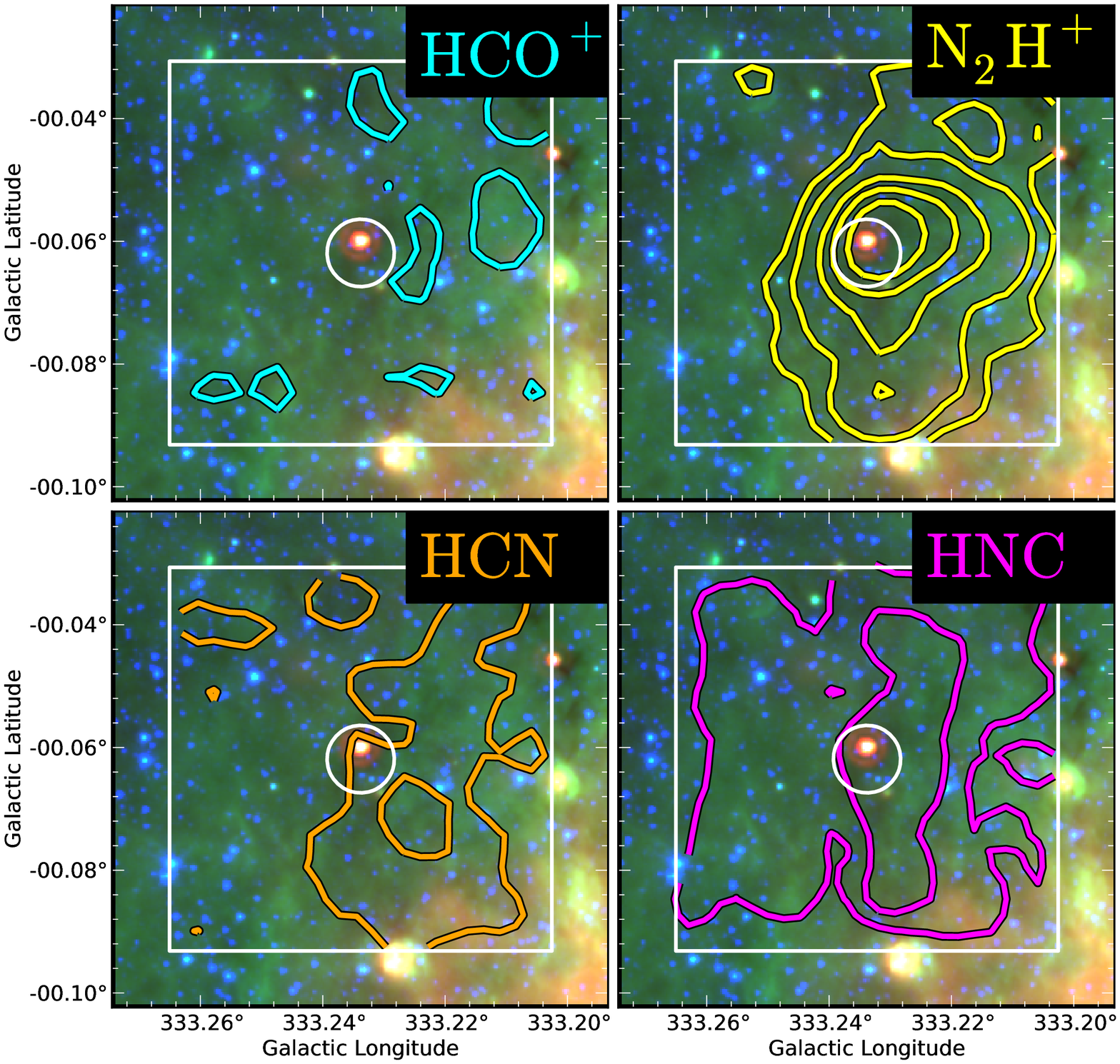}}}
%\epsscale{0.5}
%\plotone{figures/Figure13.eps}
\subfigure{\includegraphics[width=3.5in]{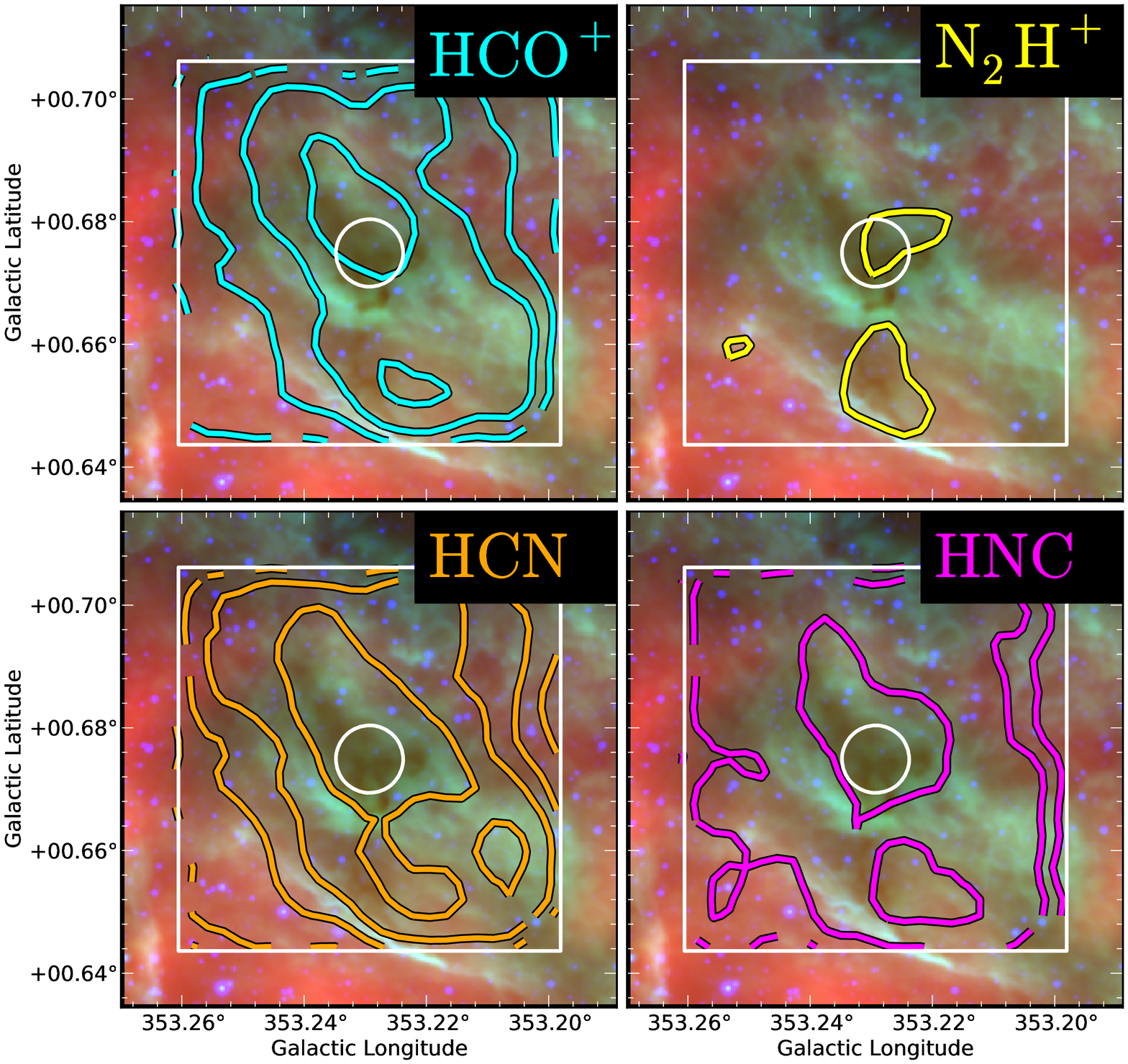}}
\caption{Examples of \nthp\ Anomalies.  Color: {\it Spitzer}/IRAC 3.6 (blue), 8.0 (green) , and {\it Spitzer}/MIPS 24 (red) $\mu$m Three-color image.  Contours: Molecular line integrated intensity image from MALT90, contours are drawn at signal-to-noise levels of 1.5, 3, 7, 11, 17. Central white circle represents the size of the Mopra beam.  White square box represents size of one Mopra map.  {\it Left}: G333.234-00.062---an example of an ``\nthp\ rich'' clump, where the integrated intensity ratio of \nthpone\ to \hcopone\ is greater than 4.0.  {\it Right}: G353.229+00.675---an example of an ``\nthp\ poor'' clump, where the integrated intensity ratio of \nthpone\ to \hcopone\ is less than 0.3.}
\label{anomaliesexamples}
%\end{center}
\end{figure*}

While measuring the \nthpone\ to \hcopone\ integrated intensity ratios [I(\nthp)/I(\hcop)] for the MALT90 sample, we identified a class of sources that we termed ``\nthp\ Anomalies'' for which the integrated intensity ratio was either unusually high or low.  Because these anomalous sources were uncommon in the MALT90 sample, they may indicate a special phase in the evolution of a high-mass star-forming clump.  ``\nthp\ rich'' sources are ones where the I(\nthp)/I(\hcop) is greater than 4.  ``\nthp\ poor'' sources are ones where the I(\nthp)/I(\hcop) is less then 0.3.  The cutoffs of 4 for rich sources and 0.3 for poor sources are arbitrary, but were chosen such that the number of poor and rich clumps were similar and set nearly an order of magnitude difference between the two extreme categories.  Figure \ref{anomaliesexamples} shows an example of a poor and rich source.

%figure7

\begin{figure*}
\begin{center}
\epsscale{0.9}
\plotone{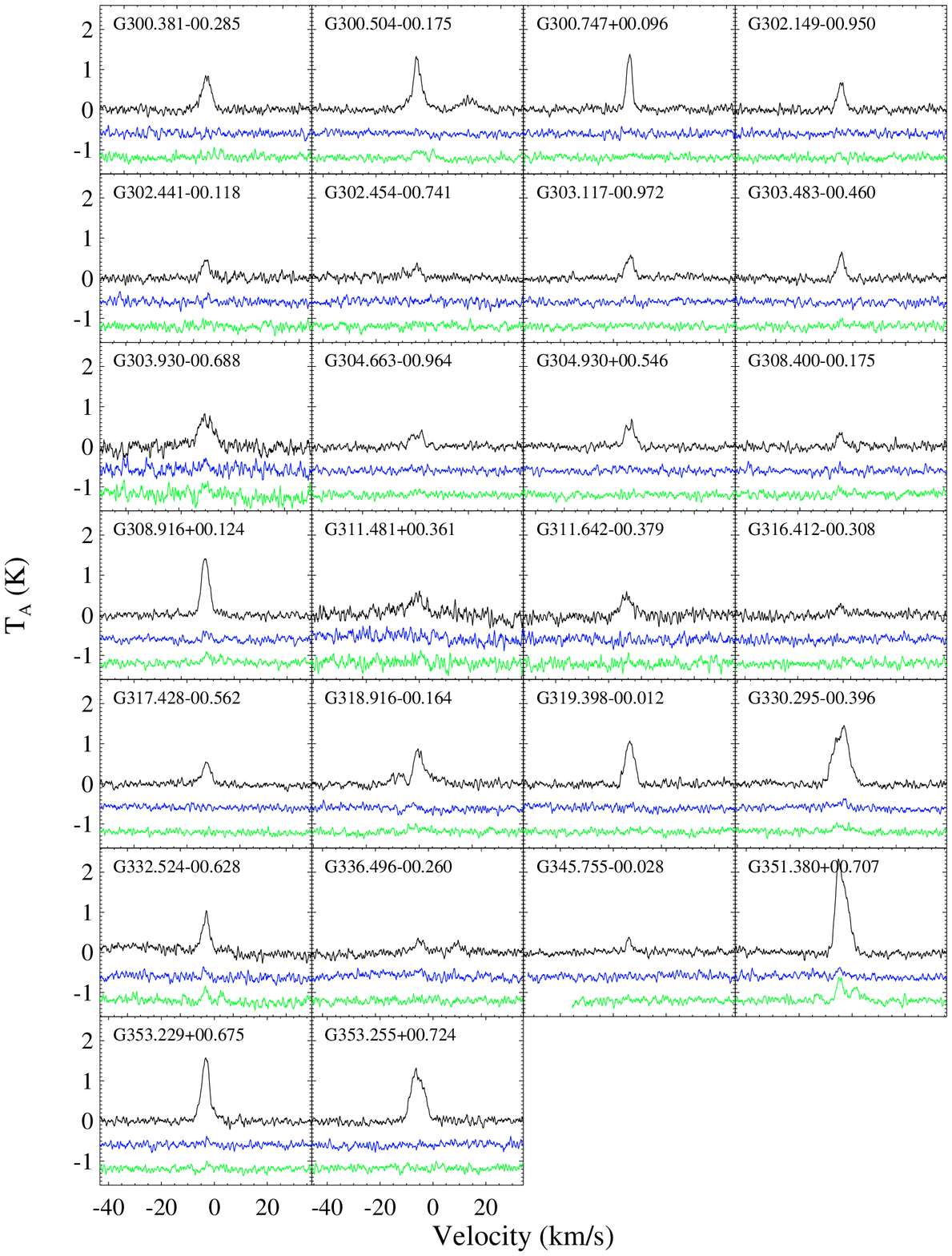}
\caption{\nthpone, \hcopone, and \htcop\ (1-0) spectra for ``\nthp poor'' sources.  \hcop\ spectra (top) are shown in black, \htcop\ spectra (middle) are in blue, and \nthp\ spectra (bottom) are in green.}
\label{n2hppoorspectra}
\end{center}
\end{figure*}

\begin{figure*}
\begin{center}
\epsscale{0.9}
\plotone{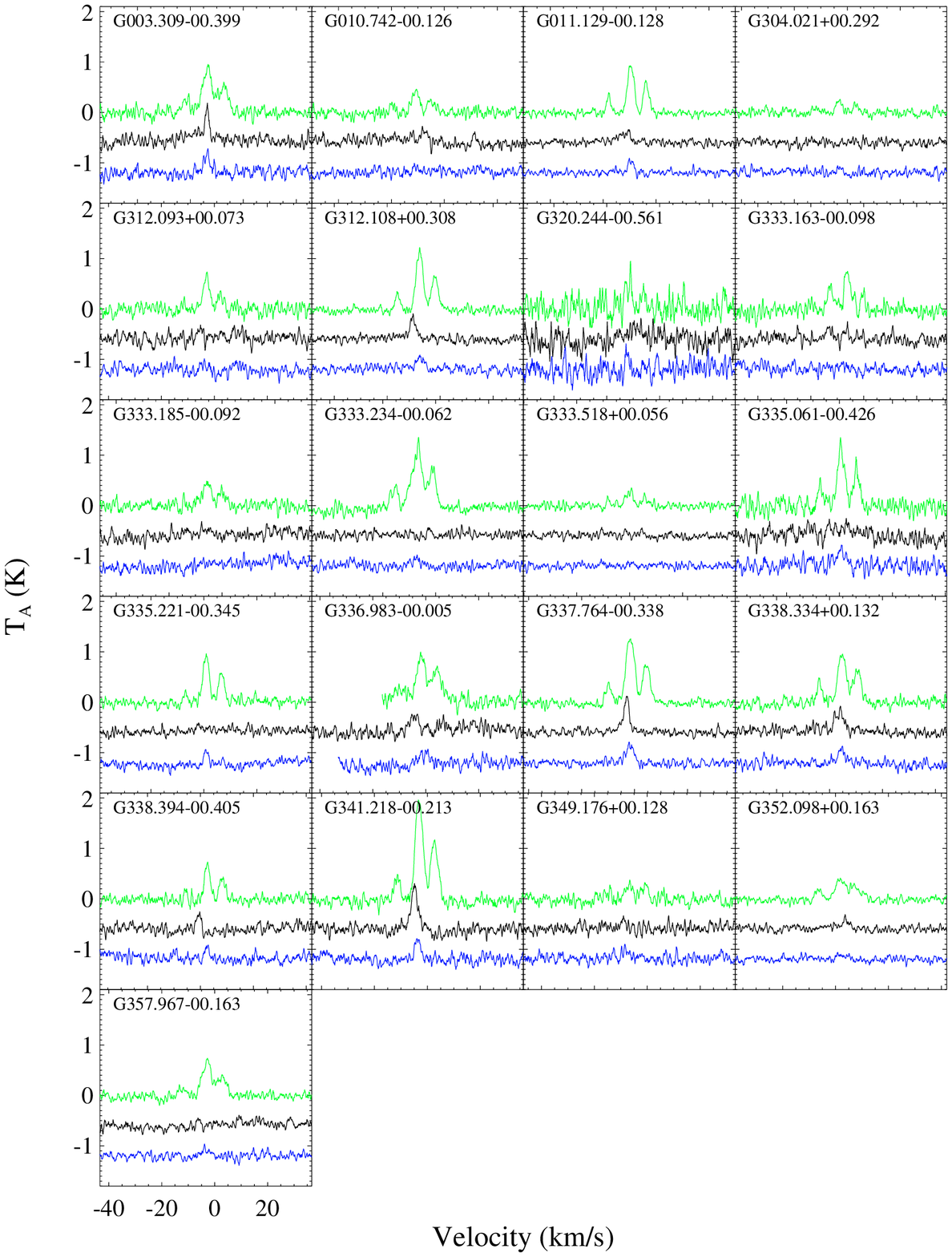}
\caption{\nthpone, \hcopone, and \htcop\ (1-0) spectra for ``\nthp\ rich'' sources.  \hcop\ spectra (middle) are shown in black, \htcop\ spectra (bottom) are in blue, and \nthp\ spectra (top) are in green.}
\label{n2hprichspectra}
\end{center}
\end{figure*}

Of the $\sim$300 sources whose I(\nthp)/I(\hcop) were calculated (see Section \ref{intensity ratio results}, 26 were categorized as \nthp\ poor and 21 as \nthp\ rich sources.  The \nthp\ poor clumps were predominently \hii/PDR regions, with 3 protostellar and 2 quiescent sources.  Of the 21 \nthp\ rich sources, 10 were \hii/PDR regions, 10 were protostellar, and one was quiescent. The \nthpone, \hcopone, and \htcop\ (1-0) spectra are shown in Figures \ref{n2hppoorspectra} and \ref{n2hprichspectra} for poor and rich sources, respectively.

%figure 8

%figure 9

Following the predictions of the behavior of \nthp\ and \hcop\ as a function of evolution, the \nthp\ poor sources should be in the later stages of evolution, which they are with the exception of the 2 quiescent sources, G332.524-00.628 and G345.755-00.028.  The properties of the \nthp\ poor sources are summarized in Table \ref{nthpdroptable}.  Based on the detection statistics of the \nthp\ poor sources determined in Section \ref{column densities and abundances}, we believe that they cannot be explained by self-absorption of \nthp.  For two \nthp\ poor sources, T$_{123-012}$ $\ge$ (5/3)$\times$T$_{112-012}$, in which case, we assume \nthp\ is optically thin (see Section \ref{optical depth}).  23 poor sources have only upper limits on the strength of the \nthp\ line, and the \nthp\ optical depth of the remaining poor source is $\sim$0.3.  We could not derive abundance ratios for six of the 26 poor sources where T$_{{\rm H^{13}CO^+ (1-0)}}$ $\ge$ T$_{{\rm HCO^+ (1-0)}}$ because we could not solve for the \hcop\ optical depth or column density (Section \ref{optical depth}).  The presence of two quiescent clumps in this sample is inconsistent with predictions of chemical evolution in star formation.  Below, we attempt to better understand these two sources by using the chemical model of \citet{Vasyunina2012}.

\begin{center}
\begin{deluxetable*}{cccccccccc}
\tabletypesize{\scriptsize}
\tablecolumns{10}
\tablecaption{Properties of ``\nthp\ poor'' Sources}
%\rotate
\tablewidth{0pt}
\tablehead{
 \colhead{Source} &
 \colhead{\underline{~I(\nthp)~}} &
 \colhead{II Ratio} &
 \colhead{Class.} &
 \colhead{II Ratio} &
 \colhead{Temp.} &
 \colhead{$\tau$$_{N_{2}H^{+}}$} &
 \colhead{$\tau$$_{HCO^{+}}$} &
 \colhead{\underline{~X(\nthp)~}} &
 \colhead{Abundance Ratio}\\
 \colhead{    } &
 \colhead{I(\hcop)} &
 \colhead{RMS} &
 \colhead{    } &
 \colhead{Detection} &
 \colhead{(K)} &
 \colhead{    } &
 \colhead{    } &
 \colhead{X(\hcop)} &
 \colhead{Detection\tablenotemark{a}}
}
\startdata
G300.381-00.285 &  0.22 &  0.07  & \hii     &        Detection      &  19.0 &   $\cdots$ &  36    &      $\cdots$  & Two Upper Limits\tablenotemark{b} \\
G300.504-00.175 &  0.18 &  0.03  & \hii     &        Detection      &  29.8 &   $\cdots$ &  17    &      $\cdots$  & Two Upper Limits \\
G300.747+00.096 &  0.06 &  0.02  & \hii     &        Upper Limit    &  22.4 &   $\cdots$ &  15    &      $\cdots$  & Two Upper Limits \\
G302.149-00.950 &  0.11 &  0.04  & \hii     &        Upper Limit    &  31.2 &   $\cdots$ &  43    &      $\cdots$  & Two Upper Limits \\
G302.441-00.118 &  0.18 &  0.06  & \hii     &        Upper Limit    &  27.3 &   $\cdots$ &  34    &        0.04   & Upper Limit \\
G302.454-00.741 &  0.23 &  0.08  & \hii     &        Upper Limit    &  $\cdots$ &   $\cdots$ &  $\cdots$ &   $\cdots$  & No HCO$^+$\tablenotemark{c} \\
G303.117-00.972 &  0.10 &  0.04  & \hii     &        Upper Limit    &  37.4 &   $\cdots$ &  50    &      $\cdots$  & Two Upper Limits \\
G303.483-00.460 &  0.30 &  0.08  & \hii     &        Detection      &  29.0 &   $\cdots$ &  53    &      $\cdots$  & Two Upper Limits \\
G303.930-00.688 &  0.22 &  0.07  & \hii     &        Detection      &  24.4 &   $\cdots$ &  24    &        0.01   & Upper Limit \\
G304.663-00.964 &  0.12 &  0.04  & Protostellar   &  Upper Limit    &  29.7 &   $\cdots$ &   $\cdots$ &  $\cdots$  & No HCO$^+$ \\
G304.930+00.546 &  0.09 &  0.03  & \hii     &        Upper Limit    &  38.0 &   $\cdots$ &  46    &      $\cdots$  & Two Upper Limits \\
G308.400-00.175 &  0.23 &  0.08  & Protostellar   &  Upper Limit    &  17.7 &   $\cdots$ &  92    &      $\cdots$  & Two Upper Limits \\
G308.916+00.124 &  0.27 &  0.03  & \hii     &        Detection      &  41.5 &   $\cdots$ &   6    &        0.03   & Upper Limit \\
G311.481+00.361 &  0.23 &  0.07  & \hii     &        Detection      &  36.8 &   $\cdots$ &   $\cdots$ &  $\cdots$  & No HCO$^+$ \\
G311.642-00.379 &  0.14 &  0.05  & \hii     &        Upper Limit    &  33.9 &   $\cdots$ &   $\cdots$ &  $\cdots$  & No HCO$^+$ \\
G316.412-00.308 &  0.29 &  0.10  & Protostellar   &  Upper Limit    &  26.3 &   $\cdots$ &   $\cdots$ &  $\cdots$  & No HCO$^+$ \\
G317.428-00.562 &  0.10 &  0.03  & \hii     &        Upper Limit    &  35.5 &   $\cdots$ &  46    &      $\cdots$  & Two Upper Limits \\
G318.916-00.164 &  0.24 &  0.04  & \hii     &        Detection      &  33.9 &   $\cdots$ &  30    &      $\cdots$  & Two Upper Limits \\
G319.398-00.012 &  0.05 &  0.02  & \hii     &        Upper Limit    &  41.5 &   $\cdots$ &  20    &      $\cdots$  & Two Upper Limits \\
G330.295-00.396 &  0.14 &  0.02  & \hii     &        Detection      &  33.1 &   $\cdots$ &   5    &        0.03   & Detection \\       
G332.524-00.628 &  0.24 &  0.06  & Quiescent &       Detection      &  17.2 &       0.34 &  15    &        0.03   & Detection \\       
G336.496-00.260 &  0.18 &  0.06  & \hii     &        Upper Limit    &  31.2 &   $\cdots$ &  38    &        0.03   & Upper Limit \\
G345.755-00.028 &  0.24 &  0.08  & Quiescent &       Upper Limit    &  17.2 &   $\cdots$ &   $\cdots$ &  $\cdots$  & No HCO$^+$ \\
G351.380+00.707 &  0.27 &  0.01  & PDR     &         Detection      &  28.9 &   $\cdots$ &   5    &        0.06   & Detection \\       
G353.229+00.675 &  0.09 &  0.02  & PDR     &         Detection      &  29.8 &   $\cdots$ &   4    &        0.07   & Upper Limit \\
G353.255+00.724 &  0.07 &  0.02  & PDR     &         Detection      &  30.0 &   $\cdots$ &  20    &      $\cdots$  & Two Upper Limits \\
\enddata
\tablenotetext{a}{The Abundance Ratio Detection refers to the detection categories of the \nthp\ and \hcop abundances in the ratio.}
\tablenotetext{b}{If both the \nthp\ and \hcop\ abundances in the ratio are upper limits, then an abundance ratio could not be calculated as it would be an unbound value.}
\tablenotetext{c}{For sources where the \hcopone\ line was not detected, the \hcop\ abundance could not be derived, and therefore, no \nthp\ to \hcop\ abundance ratio could be derived.}
\label{nthpdroptable}
\end{deluxetable*}
\end{center}

The evolutionary stages of the \nthp\ rich sources were expected to favor the early, quiescent stage.  Instead, most are \hii/PDR and protostellar sources; only one is quiescent.  For most of the rich sources, one likely explanation is that their \hcopone\ line emission is self-absorbed or optically thick, and so the I(\nthp)/I(\hcop) ratio is high because I(\hcop) is artifically low.  Since \htcop\ was detected for 17 of the 21 rich sources (albeit weakly for sources G010.742-00.126, G304.021+00.292, G333.234-00.062, and G352.098+00.163), we theorize that though the I(\hcop) was low, \hcop\ is still abundant and is self-absorbed or optically thick.  Therefore, we are left with four candidates for ``true'' \nthp\ rich sources -- G312.093+00.073, G333.163-00.098, G333.185-00.092, and G333.518+00.056, where the \htcop\ (1-0) line was not detected.  Their properties are summarized in Table \ref{nthponlytable}.  We do not derive abundance ratios for them as the \hcopone\ line was also not detected.

\begin{center}
\begin{deluxetable*}{cccccccccc}
\tabletypesize{\scriptsize}
\tablecolumns{10}
\tablecaption{Properties of ``\nthp\ rich'' Sources}
%\rotate
\tablewidth{0pt}
\tablehead{
 \colhead{Source} &
 \colhead{\underline{~I(\nthp)~}\tablenotemark{a}} &
 \colhead{II Ratio} &
 \colhead{Class.} &
 \colhead{II Ratio} &
 \colhead{Temp.} &
 \colhead{$\tau$$_{N_{2}H^{+}}$} &
 \colhead{$\tau$$_{HCO^{+}}$} &
 \colhead{\underline{~X(\nthp)~}} &
 \colhead{Abundance Ratio}\\
 \colhead{    } &
 \colhead{I(\hcop)} &
 \colhead{RMS} &
 \colhead{    } &
 \colhead{Detection} &
 \colhead{(K)} &
 \colhead{    } &
 \colhead{    } &
 \colhead{X(\hcop)} &
 \colhead{Detection}
}
\startdata
G312.093+00.073  &  9.1    &  3.1    &  \hii\  &         Lower Limit  &  24.5  &   $\cdots$  &   $\cdots$  &  $\cdots$  & No HCO$^+$ \\
G333.163-00.098  & 12.4  &  4.2  &  \hii\  &         Lower Limit  &  28.8  &   $\cdots$  &    $\cdots$  &  $\cdots$  & No HCO$^+$ \\
G333.185-00.092  &  5.5  &  1.2  &  protostellar & Detection    &  25.0  &   0.36      &   $\cdots$  &  $\cdots$  & No HCO$^+$ \\
G333.518+00.056  &  8.4  &  2.9  &  quiescent  &   Lower Limit  &  15.4  &   0.10      &   $\cdots$  &  $\cdots$  & No HCO$^+$ \\
\enddata
\tablenotetext{a}{I(\nthp)/I)\hcop) is the ratio of the integrated intensities of the \nthpone\ to \hcopone\ lines.}
\label{nthponlytable}
\end{deluxetable*}
\end{center}

For the 4 ``true'' \nthp\ rich sources, for which we theorize that the \hcopone\ and the \nthpone\ line emission is not self-absorbed, and the 2 quiescent \nthp\ poor sources, their I(\nthp)/I(\hcop) ratios should correspond to the behavior of their abundance ratios. Therefore, the unusually high I(\nthp)/I(\hcop) of the rich clumps may indicate that they are in the stage of evolution where their X(\nthp)/X(\hcop) is at a maximum.  Similarly, the low integrated intensity ratios of the two poor sources G332.524-00.628 and G345.755-00.028 may show that they are at a stage where their X(\nthp)/X(\hcop) is at a minimum.

\begin{figure*}
\begin{center}
\epsscale{0.7}
\plotone{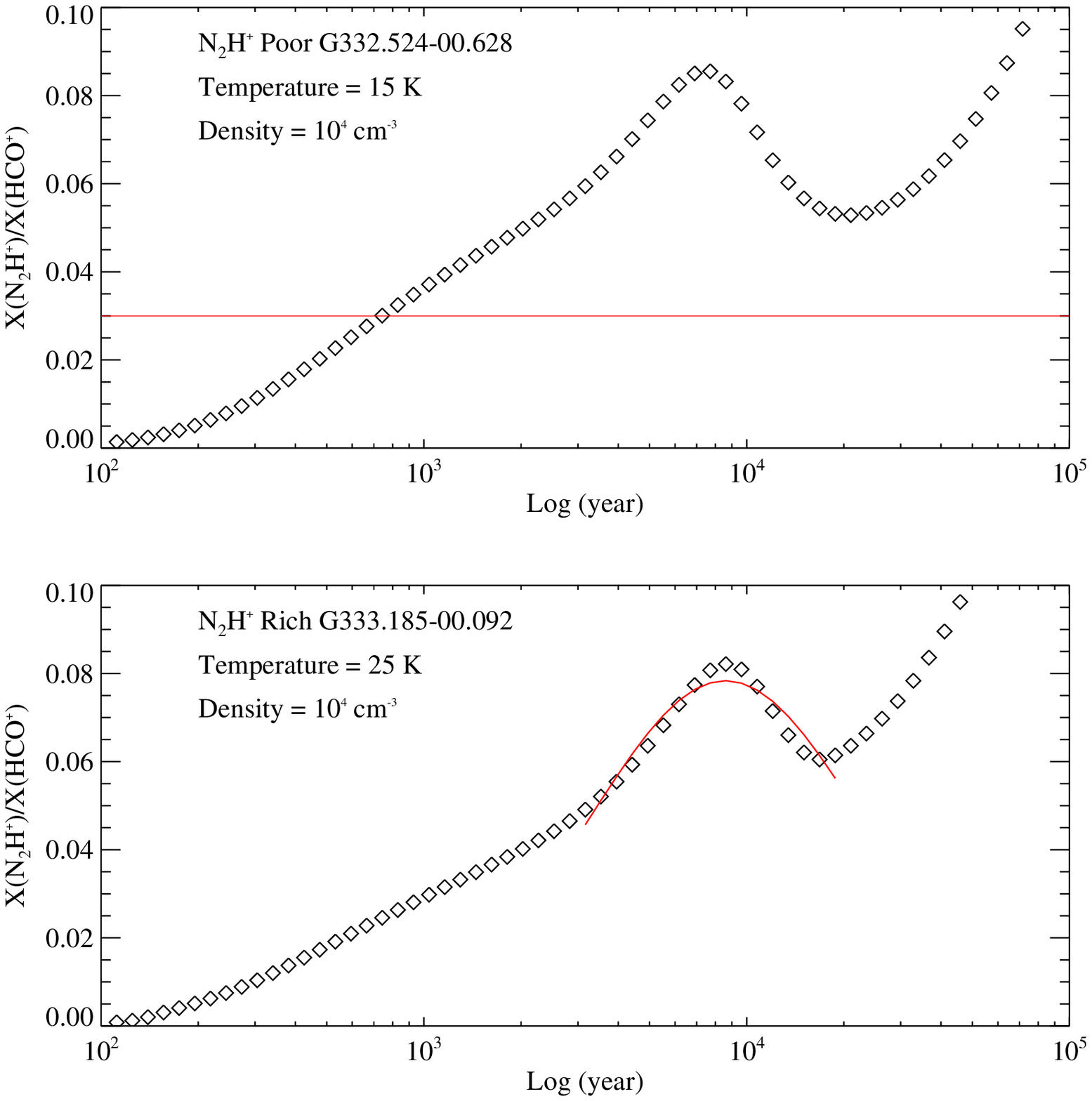}
\caption{Model X(\nthp)/X(\hcop) using as input the temperatures and estimated volume densities of \nthp\ Anomalies.  {\it Top}: The modeled X(\nthp)/X(\hcop) as a function of time corresponding to \nthp\ poor source G332.524-00.628.  The derived abundance ratio for this source is overlayed as the solid horizontal line and crosses the predicted curve near 10$^3$ years.  {\it Bottom}: The modeled X(\nthp)/X(\hcop) corresponding to \nthp\ rich source G333.185-00.092.  The peak of the ratio occurs near an age of 10$^4$ years, indicating that high-mass star-forming clumps at this age may go through a short evolutionary stage where they show an extremely high I(\nthp)/I(\hcop) ratio.  A Gaussian curve was fit to the peak to estimate the timescale of this stage, which has a FWHM $\sim$10$^4$ years.}
\label{TatianaModel}
\end{center}
\end{figure*}

To understand the anomalous \nthp\ and \hcop\ emission of the six (two poor and four rich) clumps, the \citet{Vasyunina2012} chemical evolution model was used to predict the behaviors of their \nthp\ and \hcop\ abundances.  For a given source, the model takes as input its temperature and estimated volume density and predicts how the abundances change as a function of time using gas phase, gas-grain, and grain surface chemistry.  The dust temperatures of the sources range from $\sim$15 K to 30 K, with estimated volume densities of $\sim$10$^4$ cm$^{-3}$.  The model was run using this volume density and for temperatures of 15, 25, and 30 K.  The results of the model show that the X(\nthp)/X(\hcop) increases as a function of time until it peaks near $\sim$10$^4$ years for all sources.  Figure \ref{TatianaModel} shows the predicted abundance ratio for temperatures at 15 K and 25 K, nominally corresponding to sources G332.524-00.628 and G333.185-00.092.

%figure10

There is no global minimum of abundance ratio in the model to compare to the \nthp\ poor sources.  However, we derived an abundance ratio of $\sim$0.03 for G332.524-00.628 (indicated in the figure), which corresponds to a model age of $\sim$7$\times$10$^2$ years, indicating that this is a very young clump.  However, this estimate should be taken with care because the results of chemical modeling exhibit dependence on initial conditions for very young ages of modeled objects \citep[e.g.,][]{Vasyunin2013}.  As we could not derive an abundance ratio for source G345.755-00.028,  we could not compare it to the model.

We hypothesize that the age of the peak in the model \nthp/\hcop\ abundance ratio corresponds to the age of the \nthp\ rich clumps.  Unfortunately, as the \hcopone\ line was not detected for these sources, we cannot compare the derived abundance ratios to the predicted ones.  The peak of the model abundance ratio occurs at nearly $\sim$10$^4$ years, which appears to be rather young for the two \hii\ regions, G312.093+00.073 and G333.163-00.098.  However, since the model ratio begins to rise sharply after $\sim$10$^{4.1}$ years, the ages of the two \hii\ regions may likely be $\sim$10$^5$ years, based on the model.

If extreme I(\nthp)/I(\hcop) ratios indicate special phases of evolution, then we can estimate the relative timescales of these phases from the number of \nthp\ anomalies found in the MALT90 sample.  For the \nthp\ poor sources, $\sim$25 of $\sim$300 sources were found, which would indicate this phase may last for $\sim$8\% of the total star formation timescale.  If the timescale of high-mass star formation is $\sim$10$^5$ years \citep{Zinnecker2007}, then we might expect that this phase lasts $\sim$8$\times$10$^3$ years.

If the \nthp\ rich phase also indicates a special stage in the evolutionary cycle, then based on our findings of four candidates where the \hcopone\ line does not appear to be self-absorbed or optically thick, this phase may only last for $\sim$10$^3$ years.  We also estimated the possible timescale of such a phase by measuring the apparent width of the peak in X(\nthp)/X(\hcop) predicted by the \citet{Vasyunina2012} model for G333.185-00.092.  We fit a Gaussian to the peak, whose FWHM was $\sim$10$^4$ years, which is about one order of magnitude larger than the timescale estimated by the number of \nthp\ rich sources found.  This discrepancy may be due to our selection criteria of \nthp\ anomalies, or may be because the model used does not fully represent the physical characteristics of the clumps observed.

\subsection{Can Molecular Properties be used as Chemical Clocks?}
\label{Chemical Clocks}
One goal of this study was to determine whether any chemical tracers in the evolution of massive star formation could be used to indicate the evolutionary stage.  We determined whether the correlation between \nthp\ and \hcop\ abundances, the I(HCN/HNC) and I(\nthp/\hcop) with evolutionary stage would be precise enough to establish a relationship between chemistry and evolution.

First, in studying the integrated intensity ratios, we can easily rule out I(\nthp)/I(\hcop) as a potential chemical clock.  As seen in Figure \ref{ration2hphcophcnhnchist}, the I(\nthp)/I(\hcop) shows no discernable trend as a function of evolutionary stage.  The I(HCN)/I(HNC) of the MALT90 sample shows a very weak increase as a function of evolutionary stage.  However, the distributions in each stage significantly overlap each other.  Therefore, the I(HCN)/I(HNC) is unlikely to be a precise indicator of evolution on these size scales.  Similarly, the X(\nthp)/X(\hcop) may hint at an increase as a function of evolutionary stage, but the three distributions have significant overlap and the differences in their median values are within uncertainties of the abundance ratios.

  The abundances of \nthp\ and \hcop\ suggest an increase as a function of evolution, as seen in Figure \ref{n2hphcopabundancehist}.  However, the distributions of both quantities in the three stages of quiescent, protostellar and \hii/PDR overlap, as shown by the 1-$\sigma$ intervals of the gaussian fits.

The distributions in each evolutionary stage may be indistinct for several reasons.  One major factor, stated earlier in Section \ref{comparison to predictions}, may be that the angular resolution of the observations is insufficient to isolate individual cores (size $\le$0.1 pc), and hence the chemical differences are diluted by material in the clump that is not involved in the star formation process.  For typical distances of 3 to 4 kpc for high-mass star-forming regions, and with our resolution of 38'', the spatial resolution ranges from $\sim$0.6--0.7 pc.  At these spatial resolutions, the small scale chemical variations may be lost to the overall evolution of the clump.  

Some of the overlap in the distributions may also arise from uncertainties in our evolutionary classification scheme.  Although we only include sources where the classifications were the most certain, they are nonetheless subjective, and consequently, some sources, especially distant ones, may have been classified incorrectly.

The large spread in the abundance and integrated intensity ratio distribution in each evolutionary stage of the MALT90 clumps signifies that the molecular abundances and integrated intensity ratios cannot be used as precise chemical clocks.  Given the abundance of a single source, we could not determine its evolutionary stage with certainty.  Of the four parameters though, the \nthp\ abundances show the largest variation and is the most suggestive of a chemical trend as a function of evolution.  Higher resolution data are necessary to isolate the chemistry within the dense, star-forming cores, and to determine whether the chemical clocks can be established.

\section{Conclusions}
We have observed a total of 499 high-mass star-forming clumps in 16 lines near 90 GHz as part of Year 1 of the MALT90 Survey.  In this paper, we made use of 5 molecular transitions (\nthpone, \hcopone, \htcop\ (1-0), HCN (1-0) and HNC (1-0)), and used dust continuum information from the {\it Herschel} Hi-GAL Survey.  Based on {\it Spitzer} mid-IR properties, we have classified 333 of the clumps into the evolutionary stages of quiescent, protostellar, and \hii/PDR regions.  We derived dust temperatures, H$_2$ column densities, \nthp\ and \hcop\ column densities and abundances, and \nthp/\hcop\ and HCN/HNC integrated intensity ratios for the 333 clumps.  Addressing the questions posed earlier, our main conclusions are:

%1. Based on gas masses derived from {\it Herschel} Hi-GAL dust emission maps and kinematic distance estimates from MALT90 data, the majority of MALT90 clumps are within the high-mass star-forming clump range of 10$^2$--10$^3$ M$_{\odot}$.  The majority of these clumps are likely to form one or more high-mass stars.

1. The median dust temperatures of quiescent, protostellar, and \hii/PDR regions increase with evolutionary stage, which gives confidence that our mid-IR classification scheme is a reliable indicator of evolution.

2.  Although the I(HCN)/I(HNC) ratio may suggest an increase as a function of evolutionary stage, this increase is likely not significant.

3. Contrary to the predictions of the two low-mass star formation models of \citet{Bergin1997} and \citet{Lee2004}, both \nthp\ and \hcop\ abundances increase with evolutionary stage and their abundance ratio shows a marginal increase.  This inconsistency may in part be due to the different size scales probed by the models compared to the MALT90 sample, or could hint at different chemistry in high-mass star-forming clumps.

We identified several ``\nthp\ Anomalies''-- sources where the I(\nthp)/I(\hcop) was either less than 0.3 (\nthp\ poor) or greater than 4 (\nthp\ rich), which may indicate special evolutionary phases in the high-mass star formation process.  As predicted, most of the 26 \nthp\ poor sources are all in the later stages of evolution, but the \nthp\ rich sources, which were predicted to be solely in the early stages, were often seen in the later evolutionary stages.  For all but four of these sources, this discrepancy can be explained by self-absorption of \hcop.  Using the chemical evolution model of \citet{Vasyunina2012}, we explored the two quiescent \nthp\ poor sources and the four \nthp\ rich sources that could not be explained by \hcop\ self-absorption.  
We found that the modeled \nthp\ to \hcop\ abundance ratio for all sources peak near 10$^4$ years.  Based on the number of anomalies found, the \nthp\ poor phase may last for $\sim$8\% ($\sim$10$^4$ years) of the star formation cycle while the \nthp\ rich phase may only last for $\sim$1\% ($\sim$10$^3$ years).

4. In our search for chemical clocks, the \nthp\ and \hcop\ abundances increase as a function of evolution.  The distributions in the three evolutionary stages are distinct, but have overlapping values.  To establish precise chemical clocks, high resolution observations are needed to probe individual star-forming cores housed within massive clumps.

\acknowledgements
The Mopra radio telescope is part of the Australia Telescope National Facility which is funded by the Commonwealth of Australia for operation as a National Facility managed by CSIRO.  The University of New South Wales Digital Filter Bank used for the observations with the Mopra Telescope was provided with support from the Australian Research Council.  This research has made use of the NASA/ IPAC Infrared Science Archive, which is operated by the Jet Propulsion Laboratory, California Institute of Technology, under contract with the National Aeronautics and Space Administration. NASA’s Astrophysics Data System was used to access the literature given in the references.  S.H. gratefully acknowledges funding support from NSF Grant No. AST 09-07790.  J.M.J gratefully acknowledges funding support from NSF Grant No. AST-0808001.  T. V. and A.V. wish to thank the National Science Foundation (US) for its funding of the astrochemistry program at the University of Virginia.  A.E.G. acknowledges partial support from NASA Grants NNX12AI55G and NNX10AD68G.  We thank the anonymous referee for providing valuable comments and suggestions.

\bibliographystyle{apj}
\bibliography{apjmnemonic,reflist}

%\clearpage
%%\begin{center}
%\begin{deluxetable*}{cccccc}
%\tabletypesize{\scriptsize}
%\tablecolumns{6}
%\tablecaption{MALT90 Sample of Clumps\label{sourceinfotablesample}}
%\rotate
%\tablewidth{0pt}
%\tablehead{
% \colhead{Source} &
% \colhead{RA} &
% \colhead{DEC} &
% \colhead{V$_{lsr}$} &
% \colhead{Distance\tablenotemark{a}} &
% \colhead{{\it Spitzer}} \\
% \colhead{     } &
% \colhead{     } &
% \colhead{     } &
% \colhead{(km s$^{-1}$)} &
% \colhead{(kpc)} &
% \colhead{Classification}
%}
%\startdata
%G000.253+00.016 & 17:46:09.6 & -28:42:42.6 & 36.1 & 8.3 & Quiescent \\ 
%G000.414+00.045 & 17:46:25.7 & -28:33:33.2 & 25.0 & 8.0 & Quiescent \\ 
%G003.033+00.405 & 17:51:07.1 & -26:07:45.7 & 15.7 & 4.7 & Quiescent \\ 
%G003.089+00.164 & 17:52:10.1 & -26:12:15.3 & 154.8 & 8.0 & Quiescent \\ 
%\enddata
%\tablenotetext{a}{Distances are kinematic distances as derived in Whitaker et al. 2013, in prep.}
%\tablecomments{This table is available in its entirety in a machine-readable form in the online journal. A portion is shown here for guidance regarding its form and content.}
%\end{deluxetable*}
%%\end{center}

\clearpage

\clearpage

\clearpage

\end{document}